\documentclass[twocolumn]{aastex63}

\providecommand{\e}[1]{\ensuremath{\times 10^{#1}}}	            
\providecommand{\HII}{H~{\textsc i}{\textsc i}}		        	
\providecommand{\HI}{H~{\textsc i}}	            	        	
\providecommand{\HA}{H$\alpha$}		                			
\providecommand{\R}{$R_{23}$}		                			
\newcommand{\jwst}{\emph{JWST}}		                			
\newcommand{\hst}{\emph{HST}}                                   

\received{20 December 2019}
\revised{25 February 2020}
\accepted{26 February 2020}
\submitjournal{ApJS}

\shorttitle{Dusty Sources in NGC 6822}
\shortauthors{Hirschauer et al.}
\graphicspath{{./}{figures/}}

\begin{document}

\title{Dusty Stellar Birth and Death in the Metal-Poor Galaxy NGC 6822}

\author[0000-0002-2954-8622]{Alec S.\ Hirschauer}
\affil{Space Telescope Science Institute, 3700 San Martin Drive, Baltimore, MD 21218, USA}

\author[0000-0001-6389-5639]{Laurin Gray}
\affil{Space Telescope Science Institute, 3700 San Martin Drive, Baltimore, MD 21218, USA}
\affil{Steward Observatory, University of Arizona, 933 N.\ Cherry Ave, Tucson, AZ 85721, USA}
\affil{Department of Astronomy, Indiana University, 727 E.\ 3$^{rd}$ St., Bloomington, IN 41405, USA}

\author[0000-0002-0522-3743]{Margaret Meixner}
\affil{Space Telescope Science Institute, 3700 San Martin Drive, Baltimore, MD 21218, USA}
\affil{Department of Physics \& Astronomy, Johns Hopkins University, 3400 N.\ Charles St., Baltimore, MD 21218, USA}

\author[0000-0003-4870-5547]{Olivia C.\ Jones}
\affil{UK Astronomy Technology Centre, Royal Observatory, Blackford Hill, Edinburgh, EH9 3HJ, UK}

\author[0000-0002-2996-305X]{Sundar Srinivasan}
\affil{Instituto de Radioastronom\'ia y Astrof\'isica, UNAM.\ Apdo.\ Postal 72-3 (Xangari), Morelia, Michoac\'an 58089, Michoac\'{a}n, M\'{e}xico}


\author[0000-0003-4850-9589]{Martha L.\ Boyer}
\affil{Space Telescope Science Institute, 3700 San Martin Drive, Baltimore, MD 21218, USA}

\author[0000-0001-9855-8261]{B.\ A.\ Sargent}
\affil{Space Telescope Science Institute, 3700 San Martin Drive, Baltimore, MD 21218, USA}
\affil{Department of Physics \& Astronomy, Johns Hopkins University, 3400 N.\ Charles St., Baltimore, MD 21218, USA}



\begin{abstract}

\noindent The nearby ($\sim$500 kpc) metal-poor ([Fe/H] $\approx$ --1.2; $Z$ $\approx$ 30\% $Z_{\odot}$) star-forming galaxy NGC 6822 has a metallicity similar to systems at the epoch of peak star formation.
Through identification and study of dusty and dust-producing stars, it is therefore a useful laboratory to shed light on the dust life cycle in the early Universe.
We present a catalog of sources combining near- and mid-IR photometry from the United Kingdom Infrared Telescope (UKIRT; \emph{J}, \emph{H}, and \emph{K}) and the {\it Spitzer Space Telescope} (IRAC 3.6, 4.5, 5.8, and 8.0 $\mu$m and MIPS 24 $\mu$m).
This catalog is employed to identify dusty and evolved stars in NGC 6822 utilizing three color-magnitude diagrams (CMDs).
With diagnostic CMDs covering a wavelength range spanning the near- and mid-IR, we develop color cuts using kernel density estimate (KDE) techniques to identify dust-producing evolved stars, including red supergiant (RSG) and thermally-pulsing asymptotic giant branch (TP-AGB) star candidates.
In total, we report 1,292 RSG candidates, 1,050 oxygen-rich AGB star candidates, and 560 carbon-rich AGB star candidates with high confidence in NGC 6822.
Our analysis of the AGB stars suggests a robust population inhabiting the central stellar bar of the galaxy, with a measured global stellar metallicity of [Fe/H] = -1.286 $\pm$ 0.095, consistent with previous studies.
In addition, we identify 277 young stellar object (YSO) candidates.
The detection of a large number of YSO candidates within a centrally-located, compact cluster reveals the existence of an embedded, high-mass star-formation region that has eluded previous detailed study.
Spitzer~I appears to be younger and more active than the other prominent star-forming regions in the galaxy.

\end{abstract}



\section{Introduction} 

\begin{figure*} 
\plotone{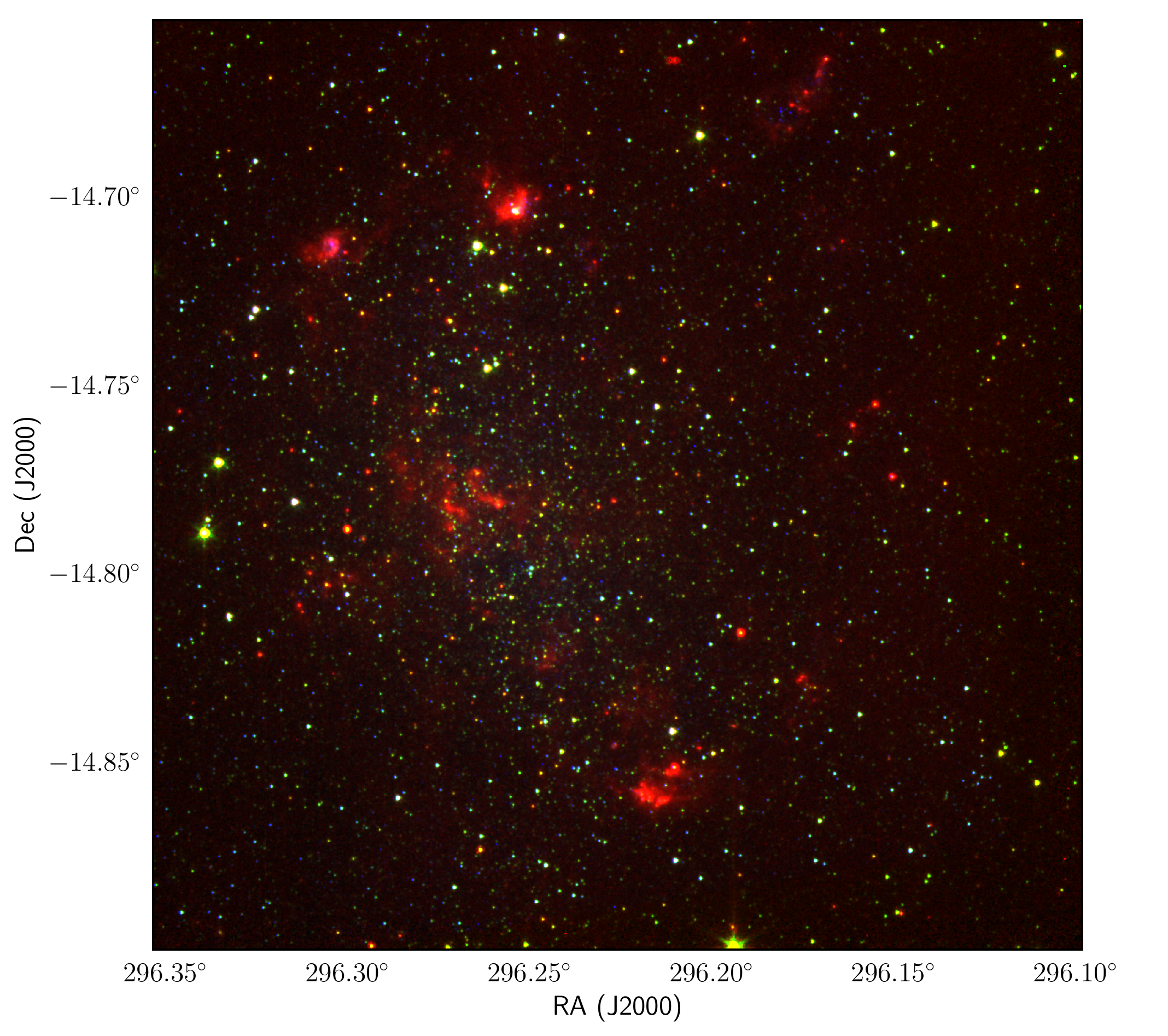}
\caption{
Composite optical and infrared (IR) image of NGC~6822.
$V$-band photometry is from the Prime Focus Direct Imager on the CTIO 4-m telescope \citep{bib:HunterElmegreen2006}, while 3.6 $\mu$m and 8.0 $\mu$m photometry is from \emph{Spitzer} IRAC \citep{bib:Khan2015}.
This nearby dwarf irregular galaxy represents an accessible analog to the star-forming systems inhabiting the Universe at higher-redshift ($z$ $\sim$ 1.5--2) useful for detailed study of star formation and chemical and dust enrichment within metal-poor environments.
NGC~6822 is home to some of the brightest star-forming region complexes in the Local Group, recognizable as clumps of red emission.
}
\label{fig:NGC6822}
\end{figure*}

\indent The nearby dwarf irregular galaxy NGC 6822 (Fig.\ \ref{fig:NGC6822}) is a well-studied member of the Local Group.
At a distance of 490 $\pm$ 40 kpc \citep{bib:Sibbons2012, bib:Sibbons2015}, its stellar populations are resolved \citep{bib:HoesselAnderson1986, bib:Gallart1994, bib:Marconi1995, bib:Gallart1996, bib:Komiyama2003, bib:deBlokWalter2006}, and with no known close companions \citep{bib:deBlokWalter2000}, it is located in an isolated environment free from gravitational interactions with other systems.
Possessing active star formation throughout its disk, NGC 6822 is home to some of the brightest giant \HII\ regions known in the local Universe \citep{bib:Hubble1925}.
Spectral abundance analyses of these nebulae have yielded low metallicities ($\sim$30\% $Z_{\odot}$; \citealp{bib:Skillman1989, bib:Lee2006}), intermediate between that of the Large and Small Magellanic Clouds (LMC/SMC; \citealp{bib:RussellDopita1992, bib:Rolleston1999, bib:Rolleston2003, bib:Lee2005b}).
The metal-poor nature and active star-formation characteristics of NGC 6822 make it an important observationally-accessible analog to the active galaxies which populated the Universe at the epoch of peak star formation ($z$ $\sim$ 1.5--2; \citealp{bib:MadauDickinson2014, bib:VanSistine2016}).
\\
\indent As a nearby star-forming galaxy, many multi-wavelength studies of NGC 6822 have been completed.
Wide-field coverage of the galaxy in optical and infrared (IR) bands have allowed for identification and classification of different stellar types.
Broadband optical data in \emph{UBVRI} was taken by \citet{bib:Massey2007}.
Warm dust emission from the interstellar medium (ISM) has been characterized with the \emph{Spitzer Space Telescope} \citep{bib:Cannon2006}, while cold dust emission was described with the \emph{Herschel Space Observatory} \citep{bib:Galametz2010}.
\citet{bib:deBlokWalter2006} studied the \HI\ distribution of NGC 6822, finding it to be unusually propagated much further beyond the main optical component of the galaxy.
The massive star population of NGC 6822 has been studied in detail by \citet{bib:Bianchi2001}.
Carbon star candidates were previously identified by \citet{bib:Letarte2002}.
Detection of RR Lyrae stars in NGC 6822 \citep{bib:Clementini2003, bib:Baldacci2004} indicate the presence of an old stellar population with an age of $\sim$11 Gyr.
The many \HII\ regions and OB associations, however, confirm that star formation is still actively ongoing.
\\
\indent Photometry in the IR traces both the beginnings and the ends of stellar lifetimes.
%
Stars in the initial stages of formation radiate strongly in IR wavelengths, as light is absorbed and re-emitted by cool accretion disks.
These young stellar objects (YSOs) possess strong IR excesses and are generally confined in proximity to regions of active star formation \citep{bib:Whitney2008, bib:Seale2009, bib:Carlson2012, bib:Jones2019}.
%
Evolved stars such as red supergiants (RSGs) and those populating the asymptotic giant branch (AGB) are among the brightest objects detected in the mid-IR.
%
RSGs and AGB stars are luminous, cool, short-lived evolutionary phases of high-mass (10 $M_{\odot}$ to 30 $M_{\odot}$) and low- to intermediate-mass (0.6 $M_{\odot}$ to 10 $M_{\odot}$) main sequence (MS) progenitors, respectively.
While RSGs trace recent star formation, AGB stars trace the old- and intermediate-age populations in galaxies \citep{bib:Blum2006}.
As AGB stars become more advanced in their evolution, convective cells and thermal pulsations drive enriched nucleosynthetic chemical products from the stellar interior toward and beyond the surface \citep{bib:HofnerOlofsson2018}.
These thermally-pulsing AGB (TP-AGB) stars can manifest as oxygen- or carbon-rich.
While the optically-thin members of this population are relatively easily separated into O- and C-rich chemistries based on their near-IR colors, the most dust-obscured TP-AGB stars (sometimes called ``extreme" AGB stars; \citealp{bib:Blum2006, bib:Boyer2011}) are not similarly discernible.
These objects are identified by their very red colors in longer wavelength bands.
As the final evolutionary phase for the vast majority of stars which have left the MS, the contribution of mass loss from RSGs and TP-AGB stars to the ISM, and the subsequent effects of this mass loss on the evolution of both the ISM and on the stars themselves, is substantial \citep{bib:HofnerOlofsson2018}.
\\
\indent Significant effort has been expended conducting recent searches for evolved stars in local galaxies with IR photometry.
For example, the DUST in Nearby Galaxies with \emph{Spitzer} survey (DUSTiNGS; \citealp{bib:Boyer2015a, bib:Boyer2015b, bib:McQuinn2017, bib:Boyer2017, bib:Goldman2019}) presented a census of dust-producing AGB stars in a sample of 50 nearby galaxies within 1.5 Mpc in 3.6 and 4.5 $\mu$m.
DUSTiNGS provided evidence that AGB stars are a major contributor of interstellar dust, even at very low metallicities.
Additionally, \emph{JHK$_{s}$} photometry from the WIYN High-resolution Infrared Camera at Kitt Peak studied RSGs and AGB stars in Sextans A and Leo A \citep{bib:Jones2018}, with results supporting the DUSTiNGS findings.
The current state of the art for dusty and evolved star identification in external galaxies is the \emph{Spitzer} Legacy Program SAGE (Surveying the Agents of Galaxy Evolution), which imaged the LMC \citep{bib:Meixner2006, bib:Blum2006, bib:Whitney2008, bib:Bernard2008} and the SMC \citep{bib:Gordon2011, bib:Boyer2011, bib:Sewilo2013}.
These programs studied the dust processes in the ISM and provided censuses of newly-formed stars, to determine the current star-formation rate (SFR; e.g., \citealp{bib:Seale2009, bib:Carlson2012}), and evolved stars, to quantify the dust-production and mass-injection rates into the ISM (e.g., \citealp{bib:Matsuura2009, bib:Boyer2012, bib:Riebel2012, bib:Srinivasan2016}).
In addition, the SAGE program generated several followup studies, including mid-IR spectroscopic data in the LMC (SAGE-Spec; \citealp{bib:Kemper2010, bib:Blum2014, bib:Jones2017a}) and SMC (SMC-Spec; \citealp{bib:Ruffle2015}), as well as observations with the \emph{Herschel} Space Telescope (HERITAGE; \citealp{bib:Meixner2010, bib:Meixner2013}) and its followup studies (e.g., \citealp{bib:Jones2015}).
\\
\indent While detailed studies of the dusty and evolved star populations within the Magellanic Clouds with SAGE are possible thanks to their close proximity and the high sensitivity afforded by \emph{Spitzer}, for a more comprehensive understanding we must extend our search to additional galaxies such as NGC 6822.
Characterization of the oxygen- and carbon-rich populations of AGB stars across the entirety of NGC 6822 was first conducted by \citet{bib:CioniHabing2005} with the William Herschel Telescope, while the distribution of O- and C-rich AGB stars across the central bar of NGC 6822 was analyzed by \citet{bib:Kang2006} using the Canada-France-Hawaii Telescope.
The distribution of O- and C-rich AGB stars was further refined by \citet{bib:Sibbons2012} using observations with the United Kingdom Infrared Telescope (UKIRT), and included an exploration of spatial variations in metallicity estimates.
Demographics of AGB stars in NGC 6822 were refined via spectroscopic study in \citet{bib:Groenewegen2009}, \citet{bib:Kacharov2012}, and \citet{bib:Sibbons2015}, employing data from the VLT and the Anglo-Australian Telescope.
Variability of the AGB star populations was studied by \citet{bib:BattinelliDemers2011} and \citet{bib:Whitelock2013} utilizing the Cerro Tololo Inter-American Observatory (CTIO) and the Japanese South-African Survey Facility (IRSF), respectively.
\\
\indent In this study, we utilize archival ground- and space-based near- and mid-IR photometric data of NGC 6822 in effort to identify RSG, TP-AGB star, and YSO candidates.
We have produced a catalog of sources compiled over a wide wavelength range that will be a useful resource for continuing and future investigations of the dusty stellar populations of NGC 6822.
These dusty sources are representative of the very early and very late stages of stellar lifetimes, providing critically-important constraint to our understanding of star formation and stellar evolution in metal-poor environments analogous to the early Universe.
We implement new techniques for developing our selection methods, emphasizing statistical rigor and a robust consideration of uncertainties.
Our resulting color cuts inform a sophisticated algorithm for categorizing the sources into several stellar classifications, including M-type stars (O-rich AGB stars and RSGs), C-type stars (C-rich AGB stars), and YSOs.
These classifications are then used to explore the spatial distribution and metallicity of the galaxy.
From this work, we prepare best practices for upcoming \emph{James Webb Space Telescope} (\jwst) guaranteed time observation (GTO) programs which will observe NGC 6822 and other metal-poor star-formation sites in the near future.
\\
\indent In \S2, we describe the archival data adopted for this study and describe the characteristics of a joined master catalog of sources in NGC 6822.
Section 3 introduces the color-magnitude diagrams (CMDs), kernel density estimate (KDE) techniques, and procedures employed in our source classifications, including the establishment of color-cut boundaries and populating the catalogs of stellar types.
In \S4 we explore the spatial distribution of the dusty and evolved sources within NGC 6822 as well as a preliminary investigation of the metallicity of the system.
We also compare our results with previous studies found in the literature.
Finally, in \S5 we summarize our findings.

\section{The Data} 

\indent This project utilized archival photometric data of NGC 6822 in the near- and mid-IR.
From the study of \citet{bib:Sibbons2012}, ground-based \emph{J}, \emph{H}, and \emph{K} photometry have been adopted from Wide Field CAMera (WFCAM) on the 3.8-m United Kingdom Infrared Telescope (UKIRT).
These data were originally obtained as part of a larger survey project of AGB stars in Local Group galaxies.
\emph{Spitzer} photometry ranging from 3.6 to 24 $\mu$m were adopted from \citet{bib:Khan2015}.
Together, the wavelength range afforded by these two datasets provides comprehensive coverage of NGC 6822 in near- and mid-IR wavelengths, emulating the coverage anticipated with \jwst.
In the following sub-sections, we describe the individual catalogs as well as the effort undertaken to join them into a single usable resource.

\subsection{UKIRT Photometry} 

\indent Near-IR photometric data has been adopted from \citet{bib:Sibbons2012}, which used the Wide Field CAMera (WFCAM) on the 3.8-m United Kingdom Infrared Telescope (UKIRT) located on Mauna Kea to image NGC 6822 in \emph{J}, \emph{H}, and \emph{K}.
This work was completed during two runs, in April 2005 and November 2006, as part of a larger survey project of Northern Hemisphere Local Group galaxies' AGB star content.
The infrared detectors of WFCAM possess a scale of 0.4$\arcsec$ per pixel over an observational area of 0.75 deg$^{2}$.
A four-tile mosaic of 3 deg$^{2}$ centered on the optical coordinates of NGC 6822 ($\alpha$ = 19$^{\text{h}}$44$^{\text{m}}$56$^{\text{s}}$, $\delta$ = --14$^{\circ}$48$\arcmin$06$\arcsec$) was obtained, with combined exposure times per pixel equalling 300 seconds in \emph{J} and 540 seconds in \emph{H} and \emph{K}.
In total, this catalog provided near-IR data for a total of 210,979 sources.
\\
\indent Data reduction followed standard steps and was completed using the WFCAM pipeline at the Institute of Astronomy in Cambridge.
Astrometric and photometric calibrations were performed based on the 2MASS point source catalog \citep{bib:Hodgkin2009, bib:Irwin2004}.
Photometric measures are based on aperture photometry, with zeropoints calibrated against 2MASS but not transformed into the 2MASS system, such that published magnitudes and colors are of the WFCAM instrumental system (transformation equations can be found in \citealt{bib:Hodgkin2009}).
Because reddening values have been found to vary across NGC 6822, with $E(B - V)$ = 0.24 mag in the outer regions and $E(B - V)$ = 0.45--0.54 mag in the center \citep{bib:Hernandez-Martinez2009b, bib:Massey1995}, no corrections were made for internal reddening.
The extinction map of \citet{bib:Schlegel1998} was utilized for implementing foreground component corrections.

\subsection{\emph{Spitzer} Photometry} 

\indent We have adopted {\it Spitzer} \citep{bib:Werner2004} mid-IR photometric data from \citet{bib:Khan2015}, which utilized archival photometric data \citep{bib:Khan2010, bib:Khan2013} originally employed to target stars of certain sub-classes with specific photometric properties.
The more recent study was designed to inventory all point sources, in effort to study seven nearby galaxies with recent star formation using the Infrared Array Camera (IRAC; \citealp{bib:Fazio2004}) at 3.6, 4.5, 5.8, and 8.0 $\mu$m and the Multiband Imaging Photometer (MIPS; \citealp{bib:Rieke2004}) at 24 $\mu$m.
This catalog contains in total 30,745 point sources from NGC 6822, created by dual-band selection of sources with $>$3$\sigma$ detections within a matching radius ($<$0.5 pixel or 0.5--1.0 pixel; 90\% of 4.5 $\mu$m sources match to a 3.6 $\mu$m source within 0.5 pixel) at both 3.6 and 4.5 $\mu$m, complemented with measurements of the 5.8, 8.0, and 24 $\mu$m bands through a combination of PSF and aperture photometry.
It was developed as a resource archive for studying mid-IR luminous sources and transients, as well as for planning observations with \jwst, which will provide much deeper observational power than {\it Spitzer}.
With a resolution nearly an order-of-magnitude better than that of {\it Spitzer} \citep{bib:Gardner2006}, data from \jwst\ will also benefit from reduced confusion.
\\
\indent This catalog is ideal for identifying dusty sources in NGC 6822.
With similar {\it Spitzer} IRAC band coverage, complemented with the 24 $\mu$m MIPS band, the \citet{bib:Khan2015} source list enables for a similar wavelength range of study to SAGE.
It provides the J2000.0 coordinates (RA and Dec) for each object, along with their Vega-calibrated apparent magnitudes ($m_{\lambda}$), the associated 1$\sigma$ uncertainties ($\sigma_{\lambda}$), and for the 3.6--8.0 $\mu$m bands, the differences between the PSF and aperture photometry magnitudes ($\delta_{\lambda}$).
\citet{bib:Khan2015} note that, for the 24 $\mu$m band photometry, the lower spatial resolution suggests limited utility, such that the relevant aperture can be contaminated by cold interstellar dust emission and objects other than the intended target.

\subsection{The Joined Catalog} 

\indent In effort to identify dusty sources in NGC 6822, we have combined the near- and mid-IR catalogs adopted from \citet{bib:Sibbons2012} and \citet{bib:Khan2015}, respectively, into a joined master catalog.
We use the CDS crossmatch utility\footnote{\texttt{http://cdsxmatch.u-strasbg.fr/xmatch}} to find the nearest \citet{bib:Sibbons2012} neighbour to each of the 30,745 sources in the \citet{bib:Khan2015} table within a 1$\arcsec$ radius.
We obtain near-IR counterparts for 14,534 sources.
This is a very small fraction of the full \citet{bib:Sibbons2012} source list, due to the large difference in survey areas between the two studies.
The near-IR counterparts include 1735 sources classified by \citet{bib:Sibbons2012} as AGB stars (1099 M-type, 636 C-type).
The 1$\arcsec$ radius allows us to recover all but nine (6 M-type, 3 C-type) of the \citet{bib:Sibbons2012} AGB candidates within the \emph{Spitzer} footprint.
Table \ref{tab:MASTERCAT} describes the columns in the joined catalog, and the full catalog is available in electronic form with this paper.
\emph{JHK} photometry adopted from \citet{bib:Sibbons2012} is presented in the WFCAM instrumental system; equations to convert to the 2MASS system are given in \citet{bib:Hodgkin2009}.

\begin{deluxetable*}{lll}
\label{tab:MASTERCAT}
\tabletypesize{\small}
\tablenum{1}
\tablewidth{0pt}
\tablecaption{Numbering, names, and description of the columns present in the master catalog, which is available as a machine-readable table.
}
\tablehead{\colhead{Column} & \colhead{Name} & \colhead{Description}}
\startdata
1 & ID\_{}MIR & Unique identifier for mid-infrared source \\
2 & RAJ2000\_{}MIR & J2000 right ascension of mid-infrared source\\
3 & DEJ2000\_{}MIR & J2000 declination of mid-infrared source\\
4 & m\_{}3.6 & IRAC 3.6 $\mu$m magnitude\\
5 & e\_{}m\_{}3.6 & Uncertainty in IRAC 3.6 $\mu$m magnitude\\
6 & m\_{}4.5 & IRAC 4.5 $\mu$m magnitude\\
7 & e\_{}m\_{}4.5 & Uncertainty in IRAC 4.5 $\mu$m magnitude\\
8 & m\_{}5.8 & IRAC 5.8 $\mu$m magnitude\\
9 & e\_{}m\_{}5.8 & Uncertainty in IRAC 5.8 $\mu$m magnitude\\
10 & m\_{}8.0 & IRAC 8.0 $\mu$m magnitude\\
11 & e\_{}m\_{}8.0 & Uncertainty in IRAC 8.0 $\mu$m magnitude\\
12 & m\_{}24 & MIPS 24 $\mu$m magnitude\\
13 & e\_{}m\_{}24 & Uncertainty in MIPS 24 $\mu$m magnitude\\
14 & ID\_{}NIR & Unique identifier for nearest-neighbor near-infrared source\\
15 & RAJ2000\_{}MIR & J2000 right ascension of nearest-neighbor near-infrared source\\
16 & DEJ2000\_{}MIR & J2000 declination of nearest-neighbor near-infrared source\\
17 & distArcsec & distance in $\arcsec$ to nearest-neighbor near-infrared source\\
18 & Jmag & {\it J}-band magnitude of nearest-neighbor near-infrared source\\
19 & e\_{}Jmag & uncertainty in {\it J}-band magnitude of nearest-neighbor near-infrared source\\
20 & Hmag & {\it H}-band magnitude of nearest-neighbor near-infrared source\\
21 & e\_{}Hmag & uncertainty in {\it H}-band magnitude of nearest-neighbor near-infrared source\\
22 & Kmag & {\it K}-band magnitude of nearest-neighbor near-infrared source\\
23 & e\_{}Kmag & uncertainty in {\it K}-band magnitude of nearest-neighbor near-infrared source\\
24 & Sibbons\_{}AGB\_{}Type & AGB classification from Sibbons et al. (2012) (`M' or `C')\\
\enddata
\tablecomments{\emph{JHK} photometry adopted from \citet{bib:Sibbons2012} is presented in the WFCAM instrumental system; equations to convert to the 2MASS system are given in \citet{bib:Hodgkin2009}.}
\end{deluxetable*}

\section{Source Classification} 

\begin{figure*} 
\plotone{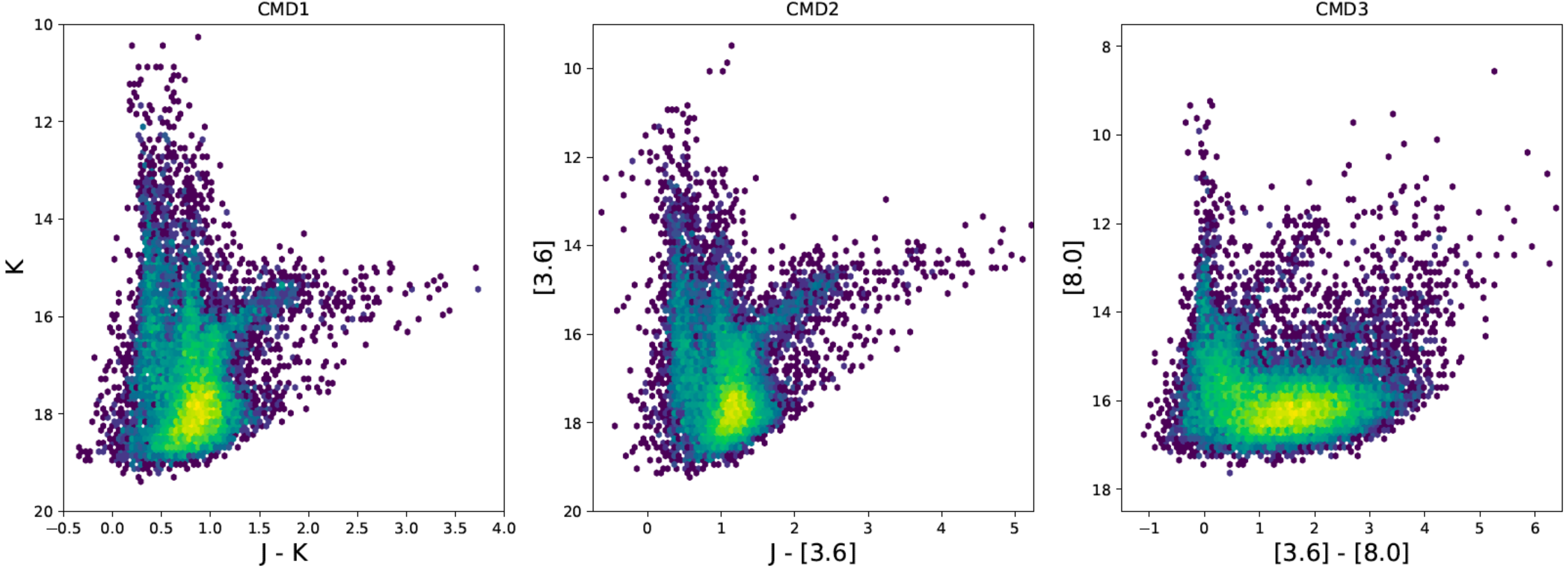}
\caption{
Hess diagrams of the three diagnostic CMDs used in this study.
Photometric data for CMD1 (\emph{left}; $K$ vs.\ $J-K$) is taken only from \citet{bib:Sibbons2012}.
Photometric data for CMD2 (\emph{center}; [3.6] vs.\ $J-[3.6]$) is taken from both \citet{bib:Sibbons2012} and \citet{bib:Khan2015}.
Photometric data for CMD3 (\emph{right}; [8.0] vs.\ $[3.6]-[8.0]$) is taken only from \citet{bib:Khan2015}.
Color cuts employing catalogs of these near- and mid-IR sources are established via KDE analyses of small, successive spans of magnitude and color.
}
\label{fig:all_hist}
\end{figure*}

\indent In order to identify and classify the dusty and evolved stars of NGC 6822, we have created three CMDs which utilize photometry from one or both archival catalogs:
$K$ vs.\ $J-K$ (hereafter ``CMD1"; Fig.\ \ref{fig:all_hist}; \emph{left}), [3.6] vs.\ $J-[3.6]$ (hereafter ``CMD2"; Fig.\ \ref{fig:all_hist}; \emph{center}), and [8.0] vs.\ $[3.6]-[8.0]$ (hereafter ``CMD3"; Fig.\ \ref{fig:all_hist}; \emph{right}).
These three CMDs were selected as representative of the wavelength ranges characteristic of each of the archival data sets (CMD1 and CMD3), as well as one which utilizes photometric data spanning both the near- and mid-IR wavelength regimes (CMD2). 
CMD1 and CMD2 show distinct features that have been described in previous studies of evolved stellar populations in nearby galaxies (e.g., \citealp{bib:Blum2006, bib:Boyer2011, bib:Jones2017b}).
Foreground objects form an almost vertical pattern in these CMDs.
The RSGs are at slightly redder colors, followed by the massive O-rich AGB stars.
The feature subsequently extending diagonally to redder colors consists of C-rich AGB stars, while beyond that are found highly-evolved AGB stars of both chemical types (the so-called ``extreme AGB" stars) and YSOs.
The emission at 8 $\mu$m is dominated by the dustiest objects, which explains the difference seen in CMD3.
At the faint end, the dusty AGB population suffers contamination from background galaxies and YSOs.
\\
\indent These three CMDs have been routinely used to devise color cuts to identify regions occupied by foreground stars, evolved stars, YSOs, and background galaxies (e.g., \citealp{bib:Boyer2011}), and the resulting classifications are found to be consistent with spectroscopic identifications (e.g., \citealp{bib:Kemper2010, bib:Jones2017b}).
Given the large sample size, the population boundaries can, in principle, be determined statistically using the observed density distribution in the CMD space.
To date, however, the color cuts have been determined largely by visual estimation alone.
\\
\indent In this paper, we present the first semi-automated selection criteria that use knowledge of the density distribution of various populations in CMD space.
Such a method has been successfully used in the literature to compute the magnitude of the tip of the red giant branch (TRGB).
\citet{bib:Cioni2000} use a histogram of the density of sources near the TRGB, finding the location where the second derivative has a maximum.
We refer the reader to Appendix A of this paper for an excellent background on the subject, as well as references to other authors who have used similar techniques.
\\
\indent As histograms come with their own problems (e.g., choice of bin sizes and bin locations; see Sections 4.8.1 and 6.1.1 in \citealp{bib:Ivezic2014}), an alternative is to smooth the observed distribution without binning.
In this paper, we use kernel density estimates (KDEs) to smooth our data, and use the local minima in these smoothed distributions to determine the boundaries between different source types in a consistent fashion.
A similar method was used by \citet{bib:Sakai1996} to determine the location of the TRGB in Sextans A.
A more recent example using smoothed distributions to compute TRGB locations is the work of \citet{bib:Freedman2019}.
\\
\indent Utilizing the three CMDs generated from our matched catalog, we identify dusty and evolved star candidates via color cuts. In this study, we present classifications for four types of dusty and evolved stars:
RSGs;
oxygen-rich AGB stars;
carbon-rich AGB stars;
and dust-enshrouded stars (including YSOs).
In addition, we identify MS stars and background galaxies, however we emphasize that our selection techniques are not optimized for these types of objects.
Our color cuts are computed by tracing the location of the local minima in kernel density estimates of the distribution in color-magnitude space. We explain this procedure in the following sections.

\subsection{Determining the Tip of the Red Giant Branch} 

\begin{figure*} 
\plotone{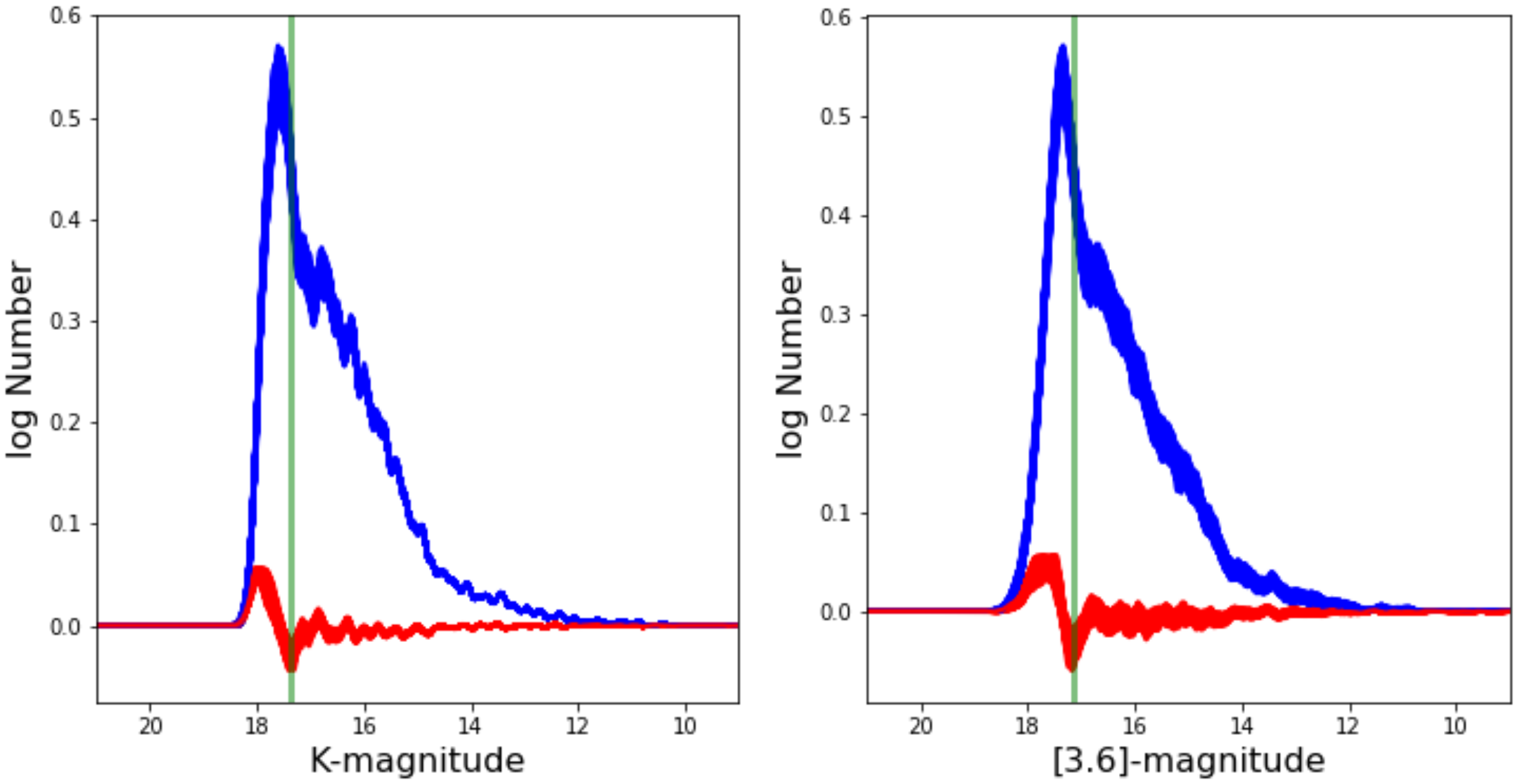}
\caption{
Our Monte Carlo + KDE method ( ``MCKDE") provides both the location of and the uncertainty associated with the TRGB for CMD1 (\emph{left}; $K$ = 17.36 $\pm$ 0.03 mag) and CMD2 (\emph{right}; [3.6] = 17.15 $\pm$ 0.06 mag), indicated with vertical green lines.
Each MCKDE routine encompasses one thousand realizations of a Gaussian kernel fit and Savistky-Golay Filter applied to the respective distribution of sources.
The location of the TRGB is determined as the point along the $x$-axis where the slope of KDE fit to the data (blue line) is steepest.
Here the absolute value of the red line (representing the magnitude of the derivative) is maximized, corresponding to the discontinuity between RGB and AGB stars.
}
\label{fig:MCKDE}
\end{figure*}

\indent Our first step in source classification is to compute the location of the TRGB.
This is an important step in the identification of dusty and evolved star candidates.
This prominent feature in the magnitude distributions of old- and intermediate-aged stellar populations corresponds to a conspicuous discontinuity between RGB and AGB populations \citep{bib:SalarisGirardi2005, bib:Sibbons2012, bib:Madore2018}.
From the Padova stellar evolution models \citep{bib:Marigo2008, bib:Marigo2013}, the study of \citet{bib:Bruzual2013} created simple stellar population models of the Magellanic Clouds, finding that $>$90\% of TP-AGB stars are brighter than the TRGB \citep{bib:Boyer2015a}.
Establishing the location of the TRGB is therefore useful as a first-order boundary used to segregate non-dusty and -evolved stars from our overall joined sample.
\\
\indent We determined the location of the TRGB using a KDE technique.
The overall distribution of the magnitudes of all sources with photometric errors less than 0.1 mag is modeled using a Gaussian kernel via the {\tt gaussian\_kde} module available in the Python {\tt SciPy} library.
In order to find the position of the TRGB in magnitude space, we compute the smoothed first derivative of the KDE by applying a Savitsky-Golay filter.
We then identify the location of the minimum in the derivative with the location of the TRGB.
In order to estimate the uncertainty associated with this estimate, we draw a large number of samples of the entire distribution using the photometric uncertainties.
The results of this Monte Carlo + KDE (MCKDE) method are shown for CMD1 and CMD2 in Fig.\ \ref{fig:MCKDE} as blue curves, the discontinuity between RGB and AGB populations is discernible as the point of steepest slope.
The value of the derivatives to the fits is represented as red curves for both CMD1 and CMD2 in Fig.\ \ref{fig:MCKDE}.
Where the absolute value of this curve is maximized corresponds to where in magnitude space lies the location of the separation between RGB and AGB stars, represented by the vertical green line.
\\
\indent For statistical robustness, the location of the TRGB was determined via Savitsky-Golay filter analysis by iterating the process one thousand times for each CMD in order to find the mean value.
These superimposed results are illustrated as Fig.\ \ref{fig:MCKDE} for both CMD1 and CMD2.
(Because of the complexity associated with the different structure of CMD3, a TRGB was not similarly determined for it;\ more detail in \S3.2.)
This Monte Carlo method of the KDE method (hereafter ``MCKDE") provides the final value of the TRGB as well as a robust estimate of its uncertainty.
For CMD1, the MCKDE method determined the TRGB at $K$ = 17.36 $\pm$ 0.04 mag (see Fig.\ \ref{fig:MCKDE}; \emph{left}), while for CMD2, the TRGB was found to be [3.6] = 17.16 $\pm$ 0.06 mag (see Fig.\ \ref{fig:MCKDE}; \emph{right}).

\subsection{Establishing Color-Cut Boundaries} 

\begin{figure*} 
\plotone{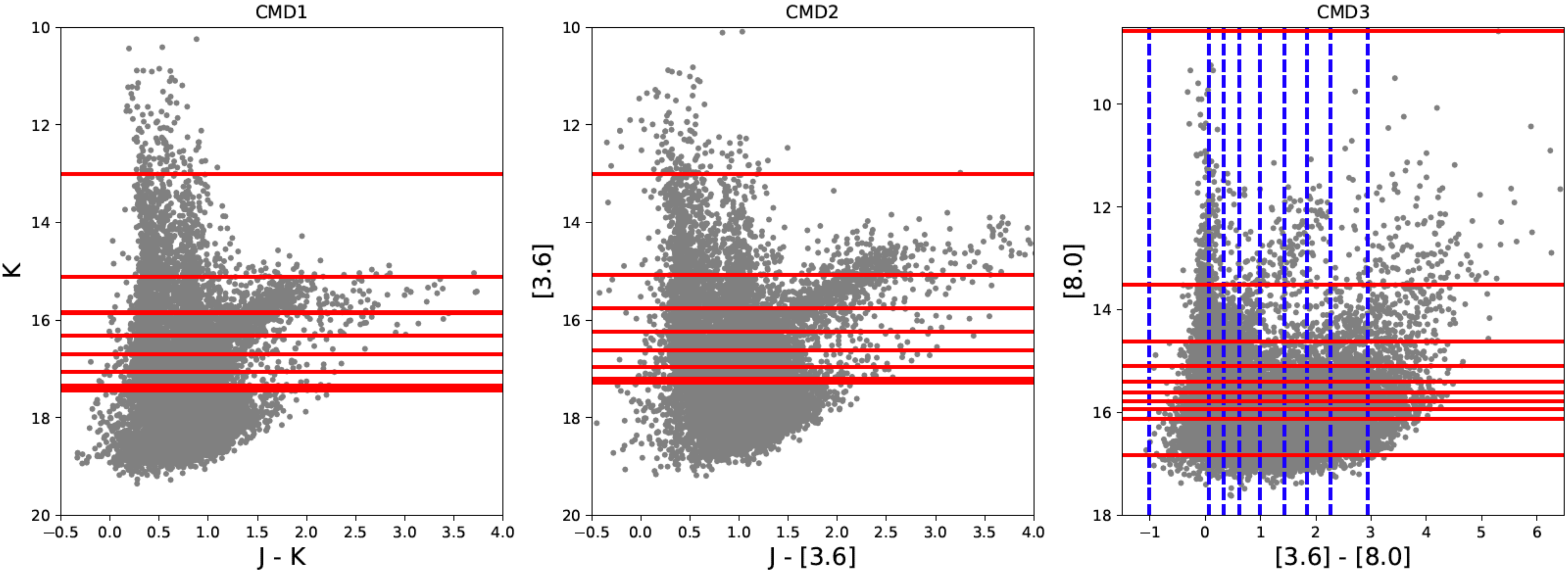}
\caption{
Successive magnitude spans utilized for each of the three diagnostic CMDs of this study.
For each of these magnitude spans, containing at least one thousand points, KDE minima analyses determine where there exists separations between distinct stellar types.
CMD1 and CMD2 (\emph{left} and \emph{center}) employ only vertically-stacked spans of successive magnitude (red lines), while CMD3 (\emph{right}) required both vertically-stacked (red solid lines) and horizontally-stacked (blue dashed lines) spans of magnitude and color, as its structure is more complicated.
Color cuts were subsequently established based on the combined results.
For CMD1 and CMD2, the TRGB is identified as the thicker red line ($K$ = 17.36 $\pm$ 0.04 and [3.6] = 17.16 $\pm$ 0.06, respectively).
}
\label{fig:all_bins}
\end{figure*}

\begin{figure*} 
\plotone{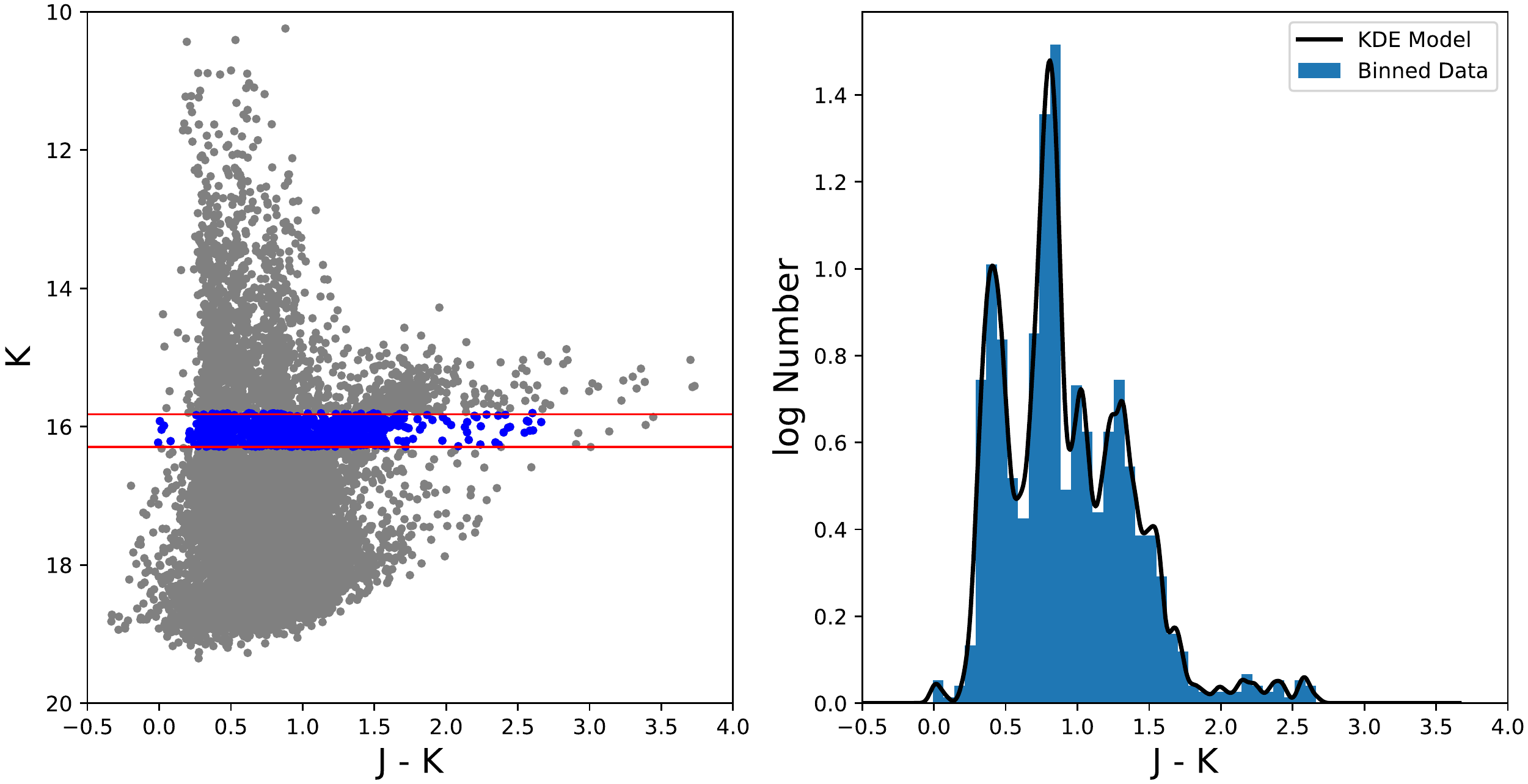}
\caption{
Example magnitude span from CMD1 (\emph{left}; $K$ = 16.316 to 15.846) and a histogram and KDE fit to the data (\emph{right}).
Red lines indicate the magnitude range for this particular span, while blue dots represent only the data within this range.
Gray points are outside the considered range.
Local minima in the KDE are representative of distinctive breaks in numbers of various stellar populations.
For each magnitude span, local minima of the central $y$-value indicate where in $J-K$-space (on the $x$-axis) a boundary between the different kinds of stars exists.
Summing the locations of these minima along the successive magnitude spans indicates where the color-cut boundaries should be placed.
This KDE local minima analysis technique was implemented for determining all color cuts.
}
\label{fig:CMDKDE_minima}
\end{figure*}

\begin{figure*} 
\plotone{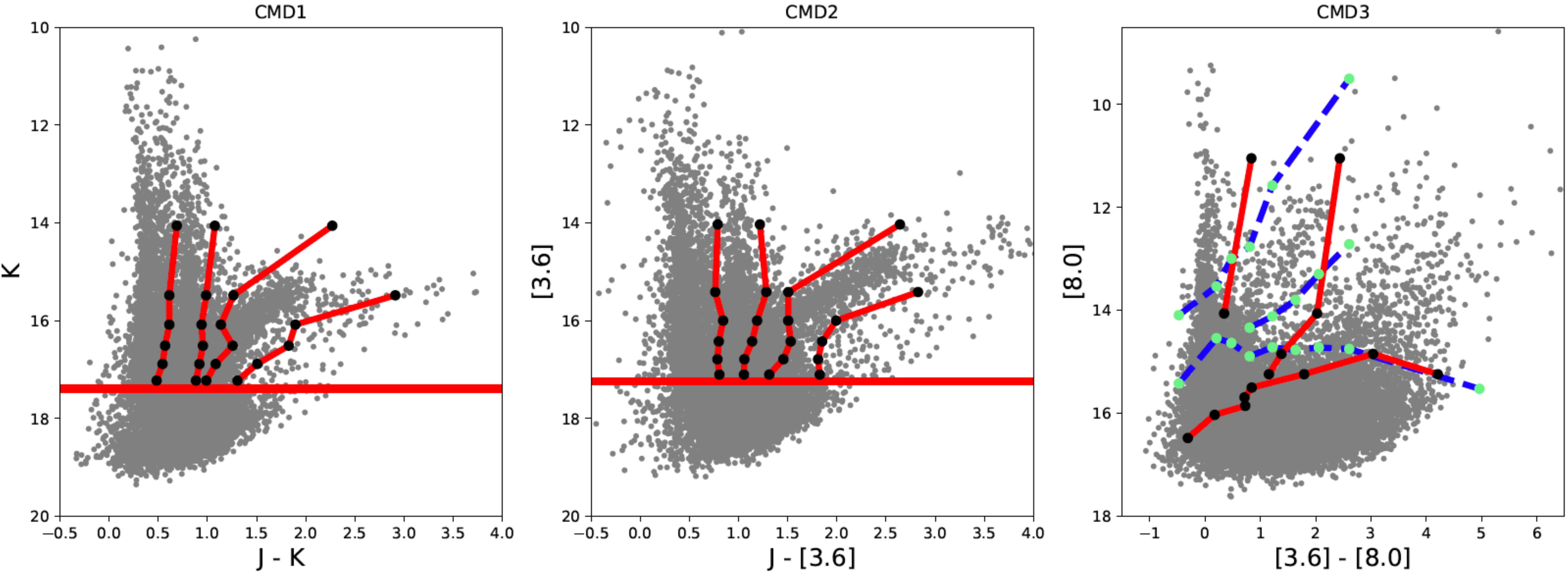}
\caption{
Boundaries separating the various stellar types in each of the three diagnostic CMDs as determined by KDE local minima analyses.
For each small span in magnitude or color, the local minima found (see Fig.\ \ref{fig:CMDKDE_minima}) indicates where stellar populations diverge.
Connecting these minima for the successive magnitude spans is then used to establish functional fits which represent the color-cut boundaries.
For CMD1 and CMD2 (\emph{left} and \emph{center}, respectively), the red lines are derived from vertically-stacked magnitude spans (the horizontal line represents the TRGB), while locations of the local minima are represented by black dots.
For CMD3 (\emph{right}), the red solid lines are derived from vertically-stacked magnitude spans while the blue dashed lines are derived from horizontally-stacked spans of color.
The local minima are represented by black and green dots, respectively.
Synthesizing both the vertically- and horizontally-stacked spans' KDE local minima into coherent color cuts required special attention to detail.
}
\label{fig:all_squiggles}
\end{figure*}

\begin{figure*} 
\plotone{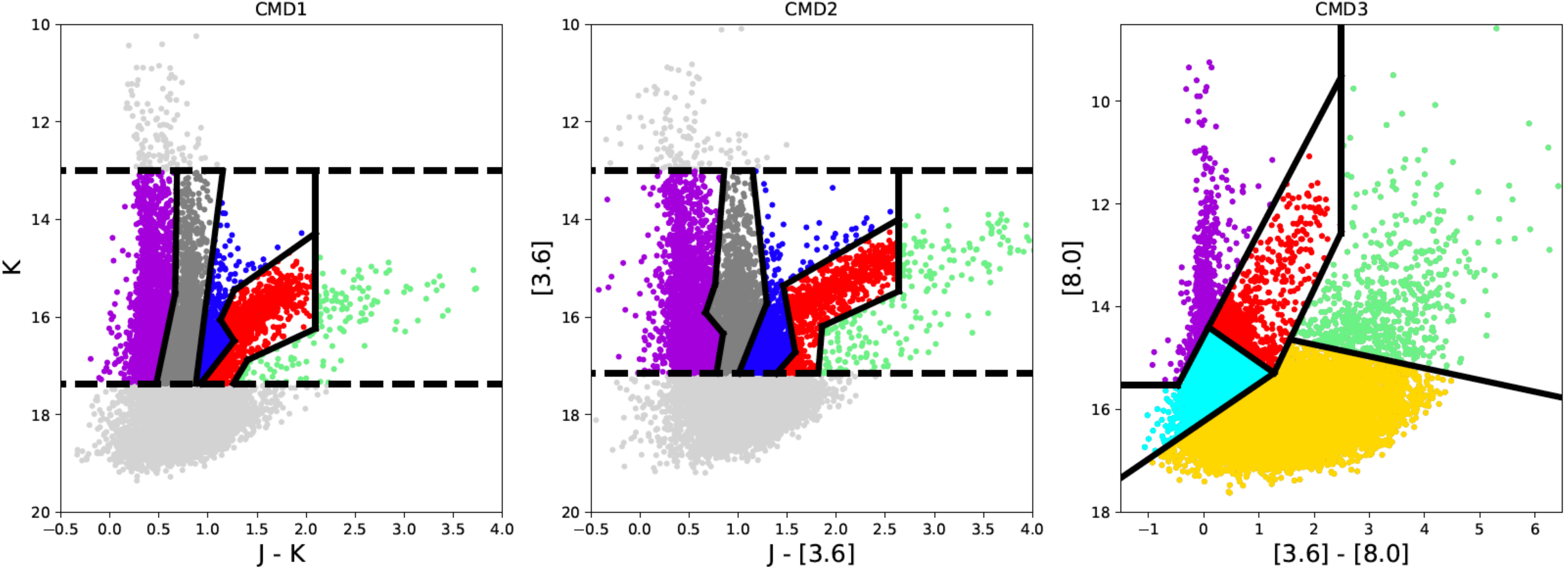}
\caption{
Color-cut boundaries (black lines) and stellar type classifications determined from each of the three diagnostic CMDs.
Violet dots are foreground/MS stars, dark gray dots are RSGs (CMD1 and CMD2 only; \emph{left} and \emph{center}, respectively), blue dots are O-rich AGB star candidates (CMD1 and CMD2 only; \emph{left} and \emph{center}, respectively), cyan points are all M-type stars, including both RSG and C-rich AGB star candidates (CMD3 only; \emph{right}), red dots are C-rich AGB star candidates, green dots are dust-enshrouded AGB stars and YSOs, gold dots represent background galaxies (CMD3 only; \emph{right}), and light gray dots represent unclassed objects.
Due to large photometric uncertainties in CMD3 below [8.0] $\approx$ 16, it is difficult to eliminate cross contamination between the cyan and gold points as CMD features become less discernible.
While the gold points are dominated by background galaxies, they may also include M-type stars and YSOs.
}
\label{fig:all_populations}
\end{figure*}

\indent The distribution of sources in each of the three CMDs presents structure representative of the underlying stellar astrophysics \citep{bib:Woods2011, bib:Ruffle2015, bib:Jones2017a}.
%
%
For a given individual source, the combination of magnitude and color reflects both the physical properties and current evolutionary phase of the star.
Such parameters are generally not obviously and easily demarcated along lines of constant magnitude or color.
The spur of points extending up and to the right from the main clustering of sources in both CMD1 and CMD2, for example, cannot be well-isolated and categorized through strictly vertical and horizontal demarcating boundary lines (e.g., \citealp{bib:NikolaevWeinberg2000, bib:WeinbergNikolaev2001, bib:CioniHabing2005, bib:Cioni2006, bib:Blum2006, bib:Boyer2011, bib:Boyer2015c}).
These clustered points in color-magnitude space clearly represent a distinct stellar type.
\\
\indent The presence of opacity differences, as well as underlying absorption features from various molecular bands and dust, have a profound effect on the photometric properties of the various stellar types.
Consequently, these populations become grouped together on the diagnostic CMDs, making it possible to distinguish them from one another.
For example, C-rich stars appear red due in part to characteristic C$_{2}$ bands between 1.15 $\mu$m and 1.3 $\mu$m \citep{bib:Groenewegen2009}, as well as HCN and C$_{2}$H$_{2}$ absorption features at 3.5 and 3.8 $\mu$m within the \emph{Spitzer} IRAC 3.6 $\mu$m band \citep{bib:Marengo2007}.
In the IRAC 8.0 $\mu$m band, silicate features such as SiO and SO$_{2}$ are prominent in O-rich evolved stars, as they are quite sensitive to the amounts of dust being produced.
For the near-IR, \citet{bib:Wright2009} presents a catalog of AGB star absorption features from molecular bands useful for modelling integrated emission for evolved stars.
In addition, \citet{bib:Blum2006} showed that, utilizing star-formation history (SFH) models of the LMC using the 2MASS CMD from \citet{bib:Cioni2006} and new stellar models from \citet{bib:Marigo2003}, C- and O-rich AGB stars were well-defined and separated in their CMDs (similar to the ones used in this study).
A strategy to group such sources under one classification which more closely resembles these structures of points is therefore to define boundaries which follow the natural shape and direction seen in the diagrams.
\\
\indent In order to define such boundaries between stellar types in color space, we implemented KDE analyses of the distribution of sources in each of the three CMDs.
With the location of the TRGB established for CMD1 and CMD2, small successive spans of magnitude incorporating at least one thousand points (for statistical robustness) were defined to progressively lower magnitude.
The vertically-stacked magnitude spans used in the analyses of CMD1 and CMD2 are sufficiently capable of separating the different stellar types (see Fig.\ \ref{fig:all_bins}; \emph{left} and \emph{center}, respectively).
Establishment of color-cut boundaries for CMD3, however, required supplemental restraint due to its more complicated structure.
Notably, the separation between some stellar types, particularly that which segregates background galaxies from dusty sources, run horizontally across the CMD.
Because this direction runs congruent with the magnitude-only spans from which the boundary lines are derived, it is difficult for the KDE procedure to determine an optimal border location.
In addition to vertically-stacked magnitude spans, we therefore additionally implemented horizontally-stacked spans in \emph{color} space, again possessing a minimum one thousand sources.
With a comparison of the demarcation lines defined by both the vertically-stacked color spans and horizontally-stacked magnitude spans, we define functional fits to the segregation boundaries of this longer-wavelength CMD (see Fig.\ \ref{fig:all_bins}; \emph{right}).
\\
\indent In each of these spans of magnitude (and/or color for CMD3), a KDE fit was established (see Fig.\ \ref{fig:CMDKDE_minima}; \emph{left}) which defined the shape of the distribution of sources (see Fig.\ \ref{fig:CMDKDE_minima}; \emph{right}).
In the left plot of this figure, within the boundaries of the CMD established by the red lines, the blue points represent only the data within the given range.
We note that some points in the magnitude ranges are not included in the KDE analysis (i.e., left as gray in the plots) as they introduced unnecessary noise to the fit that had no effect on the results.
In the right plot of this figure, a histogram of the data (blue columns) is presented as comparison with the KDE fit (black line).
Where the fit presents a local minimum is coincident with gaps in the histogram.
Locations of the local minima in each span are representative of breaks in the numbers of various stellar populations, and are therefore representative of the physically-distinct separations which exist between different stellar types.
When these local minima are joined from the contiguous successive magnitude spans, rough demarcation lines are formed (see Fig.\ \ref{fig:all_squiggles}).
Linear fits made to these points establish the functional forms of the color-cut boundaries which were used to classify the sources in NGC 6822.
\\
\indent These categorizations, illustrated in Fig.\ \ref{fig:all_populations}, include MS stars (Region 1; violet dots), RSGs (Region 2; gray dots), O-rich AGB star candidates (Region 3; blue dots), C-rich AGB star candidates (Region 4; red dots), dust-enshrouded AGB star and YSO candidates (Region 5; green dots), and background galaxies (Region 6; gold dots).
Stellar type categorizations used here should be considered as approximations drawn from a statistical sample of sources.
Furthermore, intrinsic variations of individual objects will cause some overlap.
A small degree of contamination between regions is to be expected in the diagnostic CMDs.
Final classification criteria implement region designations from multiple CMDs, which helps to reduce erroneous misidentifications (see \S3.3 for details).
%
In addition, we are not able to take into account the myriad variety of all possible stellar types.
For example, short-lived blue core helium burning sequence stars (BHeBs; also referred to as ``blue loop") present a potential source of contamination at shorter wavelengths (Regions 1 and 2; \citealp{bib:Dalcanton2009}).
Inspection of \emph{Hubble Space Telescope} (\hst) data from \citet{bib:Holtzman2006} indicate that BHeB contamination is no more than a few percent.
Finally, we note that any foreground sources are most likely Galactic MS stars, and therefore are also expected to appear within shorter-wavelength regions.
While some foreground stars will contaminate Region 2, we expect the majority to fall within Region 1.
Henceforth, objects identified within this region in any CMD will be referred to collectively as ``foreground/MS stars".

\indent For CMD1 and CMD2, the color-cut boundaries are confined within the TRGB boundaries established in \S3.1, corresponding to $K$ = 17.36 mag and [3.6] = 17.16 mag, respectively.
In addition, a second boundary for both CMD1 and CMD2 exists at $K$ and [3.6] values of 13.0 mag, whereby the number density of sources becomes too low for KDE bins to effectively constrain the source distributions.
The functional coefficients for the boundary lines separating the different stellar types in CMD1 are presented in Table \ref{tab:CMD1_boundaries} and are illustrated in Fig.\ \ref{fig:all_populations} (\emph{left}).
CMD2, which incorporates a combination of near- and mid-IR photometry, presents a distribution of sources similar in appearance to that of CMD1, but exhibits some distinct differences.
Boundary line functional coefficients which segregate the different stellar types are presented in Table \ref{tab:CMD2_boundaries} and are illustrated in Fig.\ \ref{fig:all_populations} (\emph{center}).
Finally, utilizing only mid-IR data from \emph{Spitzer} IRAC bands, CMD3 presents a structure visibly different than those seen with CMD1 and CMD2.
Rather than establishing a boundary line at the TRGB, color cuts were defined through analysis of KDE local minima analyses in both vertically- and horizontally-stacked spans of magnitude- and color-space, respectively.
This produced the complicated set of functional fits necessary to demarcate the various stellar types (see Fig.\ \ref{fig:all_populations}; \emph{right}).
Functional coefficients of the stellar type boundary lines are presented in Table \ref{tab:CMD3_boundaries}.

\begin{deluxetable*}{ccccc} 
\tablenum{2}
\label{tab:CMD1_boundaries}
\tabletypesize{\small}
\tablewidth{0pt}
\tablecaption{
Stellar type category boundary function coefficients for CMD1, following form $K$ = $m$ $\times$ ($J-K$) + $b$.
A slope of $\infty$ is representative of a vertical line.
}
\tablehead{
\colhead{Regions separated}&\colhead{slope}&\colhead{$y$-intercept}&\colhead{left extent}&\colhead{right extent}
\\
\colhead{by boundary line}&\colhead{$m$}&\colhead{$b$}&\colhead{($J-K$) =}&\colhead{($J-K$) =}
}
\startdata
1 \& 2 & -9.586 & 21.892 & 0.472 & 0.665 \\
& -128.318 & 100.790 & 0.665 & 0.684 \\
2 \& 3 & -15.880 & 31.275 & 0.876 & 1.151 \\
3 \& 4 & -2.568 & 19.735 & 0.923 & 1.265 \\
& 3.094 & 12.572 & 1.265 & 1.267 \\
& -4.380 & 20.994 & 1.127 & 1.264 \\
& -1.420 & 17.253 & 1.264 & 2.094 \\
4 \& 5 & -3.301 & 21.511 & 1.257 & 1.403 \\
& -0.934 & 18.189 & 1.403 & 2.094 \\
& $\infty$ & --- & 2.094 & 2.094 \\
\enddata
\end{deluxetable*}

\begin{deluxetable*}{ccccc} 
\tablenum{3}
\label{tab:CMD2_boundaries}
\tabletypesize{\small}
\tablewidth{0pt}
\tablecaption{
Stellar type category boundary function coefficients for CMD2, following form [3.6] = $m$ $\times$ ($J-$[3.6]) + $b$.
}
\tablehead{
\colhead{Regions separated}&\colhead{slope}&\colhead{$y$-intercept}&\colhead{left extent}&\colhead{right extent}
\\
\colhead{by boundary line}&\colhead{$m$}&\colhead{$b$}&\colhead{($J-$[3.6]) =}&\colhead{($J-$[3.6]) =}
}
\startdata
1 \& 2 & -12.033 & 26.524 & 0.779 & 0.847 \\
& 2.388 & 14.315 & 0.847 & 0.668 \\
& -5.969 & 19.898 & 0.668 & 0.766 \\
& -25.236 & 34.656 & 0.766 & 0.858 \\
2 \& 3 & -5.025 & 22.208 & 1.005 & 1.291 \\
& 19.022 & -8.849 & 1.291 & 1.149 \\
3 \& 4 & -2.287 & 20.335 & 1.390 & 1.581 \\
& 11.269 & -1.095 & 1.581 & 1.460 \\
& -1.158 & 17.046 & 1.460 & 2.637 \\
4 \& 5 & -24.317 & 61.302 & 1.816 & 1.855 \\
& -0.922 & 17.901 & 1.855 & 2.637 \\
& $\infty$ & --- & 2.637 & 2.637 \\
\enddata
\end{deluxetable*}

\begin{deluxetable*}{ccccc} 
\tablenum{4}
\label{tab:CMD3_boundaries}
\tabletypesize{\small}
\tablewidth{0pt}
\tablecaption{
Stellar type category boundary function coefficients for CMD3, following form [8.0] = $m$ $\times$ ([3.6]$-$[8.0]) + $b$.
}
\tablehead{
\colhead{Regions separated}&\colhead{slope}&\colhead{$y$-intercept}&\colhead{left extent}&\colhead{right extent}
\\
\colhead{by boundary line}&\colhead{$m$}&\colhead{$b$}&\colhead{([3.6]$-$[8.0]) =}&\colhead{([3.6]$-$[8.0]) =}
}
\startdata
1 \& 2/3 & 0 & 15.529 & $\infty$ & -0.475 \\
& -2.029 & 14.565 & -0.475 & 2.485 \\
& $\infty$ & --- & 9.522 & 9.522 \\
2 \& 3 & 0.749 & 14.332 & 0.084 & 1.273 \\
2 \& 6 & -0.735 & 16.226 & $\infty$ & 1.273 \\
3 \& 5/6 & -2.251 & 18.151 & 1.273 & 2.485 \\
& $\infty$ & --- & 12.556 & 12.556 \\
5 \& 6 & 0.229 & 14.277 & 1.562 & $\infty$ \\
\enddata
\end{deluxetable*}

\indent The stellar classification categories established from our analyses are presented as Table \ref{tab:MASTER_class}.
These columns represent an addendum to the master joined catalog (Table \ref{tab:MASTERCAT}) and are included as part of the downloadable dataset.
The full overall dataset is available in electronic, machine-readable format.
Included are an object ID number, the RA and Dec coordinates from \citet{bib:Sibbons2012}, and the coordinate disparity between the UKIRT and \emph{Spitzer} photometry measured in arcseconds, where applicable.
The \citet{bib:Khan2015} data (\emph{Spitzer} [3.6], [4.5], [5.8], [8.0], and [24]) presents Vega-calibrated apparent magnitudes ($m_{\lambda}$) and the associated 1$\sigma$ uncertainties ($\sigma_{\lambda}$)
For the 5.8, 8.0, and 24 $\mu$m bands, a 1$\sigma$ uncertainty of ``-9999" corresponds to the photometric flux being a 3$\sigma$ upper limit.
\citet{bib:Sibbons2012} data (UKIRT \emph{J}, \emph{H}, and \emph{K}) were calibrated astrometrically and photometrically to the 2MASS point source catalog \citep{bib:Hodgkin2009, bib:Irwin2004}.
Photometric measurements are based on aperture photometry, using zeropoints calibrated against (but not transformed into) the 2MASS system.
\\
\indent Finally, for each source in the master joined catalog we include an identifier value signifying the color-cut region within which the source is located for each of the three diagnostic color-magnitude diagrams (CMDs; see \S3.2).
As with the CMD classification regions of Fig.\ \ref{fig:all_populations}, ``1" signifies that the source is located in the color-cut region of foreground/MS stars.
``2" represents RSG candidates, while sources marked ``3" belong to O-rich AGB star candidates.
C-rich AGB star candidates are notated with ``4", while ``5" are dust-enshrouded sources which include dusty AGB stars and young stellar objects (YSOs).
Finally, sources marked ``6" are background galaxy candidates, and any object denoted as ``-9999" were not classified in the given CMD (did not inhabit an available color-cut boundary region).
The final column, ``TYPE", gives the source classification as defined by the procedure outlined in \S3.3.
``RELIABLE" classifications are denoted as one of the following descriptors:
``RSG", ``O-rich", ``C-rich", or ``YSO".
Furthermore, ``CANDIDATE" sources are indicated as ``RSG?", ``O-rich?", ``C-rich?", or ``Dusty?".
In addition, we note ``Star", ``Galaxy", and ``Unclassifiable" sources based on the categorizations provided by our CMD analyses, though we note that our selection criteria are not specified for these kinds of objects.
Finally, sources for which a classification was not established are presented as ``Classless!".

\begin{deluxetable}{lll} 
\tablenum{5}
\label{tab:MASTER_class}
\tabletypesize{\small}
\tablewidth{0pt}
\tablecaption{
Addendum to master catalog (Table \ref{tab:MASTERCAT}) with CMD categorizations and final type classifications.
}
\tablehead{
\colhead{Column}&\colhead{Name}&\colhead{Description}
}
\startdata
25 & CMD1 & Region Categorization for CMD1 \\
26 & CMD2 & Region Categorization for CMD2 \\
27 & CMD3 & Region Categorization for CMD3 \\
28 & TYPE & Final Type Classification \\
\enddata
\end{deluxetable}

\subsection{Catalogs of Stellar Types} 

\begin{deluxetable}{ccc} 
\tablenum{6}
\label{tab:CMDclassifications}
\tabletypesize{\small}
\tablewidth{0pt}
\tablecaption{
CMD color cut initial source classifications.
}
\tablehead{
\colhead{CMD}&\colhead{Stellar Type}&\colhead{Number of Sources}
}
\startdata
$K$ vs.\ $J-K$ & RSG & 2,075 \\
$K$ vs.\ $J-K$ & O-rich AGB & 1,002 \\
$K$ vs.\ $J-K$ & C-rich AGB & 907 \\
$K$ vs.\ $J-K$ & Dust-enshrouded & 144 \\
{[}3.6] vs.\ $J-[3.6]$ & RSG & 1,571 \\
{[}3.6] vs.\ $J-[3.6]$ & O-rich AGB & 1,545 \\
{[}3.6] vs.\ $J-[3.6]$ & C-rich AGB & 713 \\
{[}3.6] vs.\ $J-[3.6]$ & Dust-enshrouded & 230 \\
{[}8.0] vs.\ $[3.6]-[8.0]$ & M-type stars & 2,949 \\
{[}8.0] vs.\ $[3.6]-[8.0]$ & C-rich AGB & 674 \\
{[}8.0] vs.\ $[3.6]-[8.0]$ & Dust-enshrouded & 630 \\
\enddata
\end{deluxetable}

\indent From the established color-cut boundaries, we have designated a type classification for every source in each of the three CMDs (see Fig.\ \ref{fig:all_populations}).
Guided by the KDE analyses described in the previous section, the demarcation lines indicate whether a given source is:\
a foreground/MS star (Region 1; violet points); an RSG candidate (Region 2 (CMD1 and CMD2 only); gray points); an O-rich AGB star candidates (Region 3 (CMD1 and CMD2 only); blue points); a C-rich AGB star candidate (Region 4; red points); a dust-enshrouded star, typical of extreme AGB star candidates and YSOs (Region 5; green points); or a background galaxy (Region 6 (CMD3 only); gold points).
For CMD3 only as well, Region 2 (cyan points) corresponds to all M-type stars, which includes both RSGs and O-rich AGB star candidates.
If a source fails to be categorized within any of the regions described above, it is left as classless (represented as faint gray points in CMD1 and CMD2 of Fig.\ \ref{fig:all_populations}).
\\
\indent The stellar type classification demographics derived from each of the three CMDs are presented in Table \ref{tab:CMDclassifications}.
Remarkably, the stellar type demographics from the analyses of each CMD produce numbers which are generally self consistent.
This is true despite the KDE analyses for each of the three CMDs being completed entirely independent from one another.
The only obvious exception is the elevated number of dust-enshrouded sources identified in CMD3 as compared to CMD1 and CMD2.
This disparity is attributable to the longer wavelengths of the photometric bands employed, resulting in an increased sensitivity to the dusty objects which primarily emit in this range.
\\
\indent Utilizing the stellar type categorizations as determined by each CMD individually, an overall classification for every source in the master catalog was determined.
These compared the categorization results from each CMD to those from each of the other two CMDs in effort to reach a consensus.
High-confidence classifications by robust combinations of individual CMD results were classified as ``RELIABLE", while lower-confidence classifications were marked as ``CANDIDATE".
From a total overall catalog of $N$ = 30,745 sources, $n$ = 3,179 (10.3\%) have been classified as a ``RELIABLE" dusty or evolved star.
Our selection criteria stipulate relative consistency in stellar type categorization across the three CMDs.
Because some differences in categorization can result as a consequence of the different sensitivities afforded to the various wavelength regimes (e.g., CMD1 can better identify foreground stars, while CMD3 is better for classifying very dusty sources), such small inconsistencies will not disqualify a source from a ``RELIABLE" label.
\\
\indent A source is considered a ``RELIABLE" M-type star candidate when it is classified as an RSG or an O-rich AGB star in CMD1 and CMD2, and so long as it is not categorized as a dust-enshrouded source in CMD3.
Of the M-type stars, RSGs are those which inhabit Region 2 in both CMD1 and CMD2, while O-rich AGB stars are those which inhabit Region 3 in both CMD1 and CMD2, or else is categorized as Region 3 in one of these two CMDs and as Region 2 in the other.
Categorization within Region 3 in at least one of CMD1 or CMD2 favors a source to be an O-rich AGB star rather than an RSG.
%
%
Previous study of the LMC employing CMDs equivalent to our CMD1 \citep{bib:NikolaevWeinberg2000} and CMD2 \citep{bib:Blum2006} identified features blue-ward of the O-rich AGB star sequence, categorizing the relevant sources as RSGs.
These classifications were then confirmed through spectroscopic cross-correlation.
\citet{bib:Boyer2011} extended this technique to the SMC to assess a distinction between RSGs and O-rich AGB stars, which was later revisited by \citet{bib:Yang2018} and \citet{bib:Yang2019}.
With theoretical support from the modeling of \citet{bib:DellAgli2015}, we have adopted this prior proven technique for understanding the distinction between RSGs and O-rich AGB stars in CMDs applied to other galaxies.
\\
\indent In total, $n$ = 2,342 sources have been classified as ``RELIABLE" M-type star candidates, with $n$ = 1,292 of these being designated as RSGs and $n$ = 1,050 as O-rich AGB stars.
``RELIABLE" C-type star candidates are those which are categorized as C-rich in at least two of the three diagnostic CMDs, and are additionally never categorized as dust-enshrouded in any of the three CMDs.
There are $n$ = 560 sources classified as ``RELIABLE" C-type star candidates.
A source is considered a ``RELIABLE" YSO candidate only when it is categorized as a dust-enshrouded object in each of the three CMDs.
In addition, because YSOs are definitively restricted to regions of recent star-formation, which are found in clusters, we have included a secondary requirement that a YSO is categorized as ``RELIABLE" only if it found within one of seven star-formation regions where the number density of YSO candidate sources is highest (e.g., \citealp{bib:Jones2019}).
This yields a final total of ``RELIABLE" YSO candidates of $n$ = 277.
The classification criteria for these sources is presented in Table \ref{tab:RELIABLE}).
\\
\indent Other types of sources, including background galaxies and foreground/MS stars, were similarly classified based on their categorizations from the three diagnostic CMDs.
Because our methods are not optimized for such sources, we refrain from detailed explanations of these objects' classifications.
Rather, we note that such sources were identified outside of the RSG, AGB star, and YSO candidate color cuts.
Finally, sources which are determined to be reliably unclassifiable are those which do not receive categorization into any of the possible stellar types in any of the diagnostic CMDs.
\\
\indent A further $n$ = 260 (0.8\%) objects have been classified as ``CANDIDATE" sources, indicating that the confidence level in their combined stellar-type categorizations is lower than for the ``RELIABLE" targets.
This is typically due to more pronounced inconsistencies in classification defined across the three CMDs than was allowable for the ``RELIABLE" targets, while retaining photometric characteristics consistent with a particular stellar type.
While very few such ``CANDIDATE" sources populate our catalog as compared to ``RELIABLE" sources, we felt their inclusion was warranted as a preliminary means of classification, with expectation for more sophisticated techniques applied later.
The combination of diagnostic CMD categorizations for ``CANDIDATE" sources follow similar structure to those of ``RELIABLE" sources.
These selection criteria are summarized in Table \ref{tab:CANDIDATES}.
In total, our catalog of ``CANDIDATE" objects includes $n$ = 65 RSGs, $n$ = 59 O-rich AGB stars, $n$ = 27 C-rich AGB stars, and $n$ = 109 extremely-dusty objects.
We note that a source that would qualify as a ``RELIABLE" YSO but lies outside of the star-forming regions is thus classified as ``CANDIDATE" C-type star.
These dust-enshrouded sources, most predominantly selected from the longer-wavelength color cuts of CMD3, include both YSOs and perhaps a rare class of the dustiest extreme AGB stars (``x-AGB" stars; see \citealp{bib:Blum2006, bib:Boyer2011, bib:Boyer2015a}).
These x-AGB stars have been found to produce an extremely large proportion of dust for their small numbers in metal-poor star-forming environments.
\\
\indent Finally, the remaining $n$ = 1,660 (5.4\%) sources are left as ``CLASSLESS".
These objects' photometric properties provided no obvious identity based on categorical consistency within the three CMDs and were therefore left as unclassified.
Comprising a relatively small minority of the overall sample, we expect future categorization with more sophisticated identification techniques to shed light on these sources.
The demographic results of the ``RELIABLE" and ``CANDIDATE" catalogs are summarized in Table \ref{tab:catalog_demographics}, along with all ``OTHER" sources.

\begin{deluxetable*}{cccccc} 
\tablenum{7}
\label{tab:RELIABLE}
\tabletypesize{\small}
\tablewidth{0pt}
\tablecaption{
Categorization region combinations amalgamated from the CMD classifications for ``RELIABLE" RSGs, O-rich AGB stars, C-rich AGB stars and YSOs.
Note that YSOs require the additional criterion that these sources fall within one of the seven star-forming regions demarcated in Fig.\ \ref{fig:posterpic}.
}
\tablehead{
\colhead{Type}&\colhead{CMD1 region}&\colhead{}&\colhead{CMD2 region}&\colhead{}&\colhead{CMD3 region}
\\
\colhead{}&\colhead{$K$ vs.\ $J-K$}&\colhead{}&\colhead{[3.6] vs.\ $J-$[3.6]}&\colhead{}&\colhead{[8.0] vs.\ [3.6]$-$[8.0]}
}
\startdata
RSG:\ & 2 & \& & 2 & \& & \textsc{not} 5 \\
O-rich AGB star:\ & 2 & \& & 3 & \& & \textsc{not} 5 \\
& 3 & \& & 2 & \& & \textsc{not} 5 \\
& 3 & \& & 3 & \& & \textsc{not} 5 \\
C-rich AGB star:\ & 4 & \& & 4 & \& & \textsc{not} 5 \\
& 4 & \& & \textsc{not} 5 & \& & 4 \\
& \textsc{not} 5 & \& & 4 & \& & 4 \\
YSO:\ & 5 & \& & \ldots & \& & \ldots \\
& \ldots & \& & 5 & \& & \ldots \\
& \ldots & \& & \ldots & \& & 5 \\
\enddata
\end{deluxetable*}

\begin{deluxetable*}{cccccc} 
\tablenum{8}
\label{tab:CANDIDATES}
\tabletypesize{\small}
\tablewidth{0pt}
\tablecaption{
Categorization region combinations amalgamated from the CMD classifications for ``CANDIDATE" O-rich AGB stars, C-rich AGB stars, and extremely-dusty objects.
}
\tablehead{
\colhead{Type}&\colhead{CMD1 region}&\colhead{}&\colhead{CMD2 region}&\colhead{}&\colhead{CMD3 region}
\\
\colhead{}&\colhead{$K$ vs.\ $J-K$}&\colhead{}&\colhead{[3.6] vs.\ $J-$[3.6]}&\colhead{}&\colhead{[8.0] vs.\ [3.6]$-$[8.0]}
}
\startdata
RSG:\ & 2 & \& & 1 & \& & 1 \\
O-rich AGB star:\ & 3 & \& & 4 & \& & 2 \\
C-rich AGB star:\ & 4 & \& & 5 & \& & 4 \\
extremely dusty:\ & 5 & \& & 5 & \& & 4 \\
\enddata
\end{deluxetable*}

\begin{deluxetable*}{cccccc} 
\tablenum{9}
\label{tab:OTHER}
\tabletypesize{\small}
\tablewidth{0pt}
\tablecaption{
Categorization region combinations amalgamated from the CMD classifications for background galaxies and foreground/MS stars.
}
\tablehead{
\colhead{Type}&\colhead{CMD1 region}&\colhead{}&\colhead{CMD2 region}&\colhead{}&\colhead{CMD3 region}
\\
\colhead{}&\colhead{$K$ vs.\ $J-K$}&\colhead{}&\colhead{[3.6] vs.\ $J-$[3.6]}&\colhead{}&\colhead{[8.0] vs.\ [3.6]$-$[8.0]}
}
\startdata
Background galaxy: & 1 & \& & 1 & \& & 6 \\
& 1 & \& & \ldots & \& & 6 \\
& \ldots & \& & 1 & \& & 6 \\
& \ldots & \& & \ldots & \& & 6 \\
Foreground/MS star: & 1 & \& & 1 & \& & 1 \\
& 1 & \& & 1 & \& & 2 \\
& 1 & \& & 1 & \& & 5 \\
& 1 & \& & 1 & \& & \ldots \\
& 1 & \& & 2 & \& & 2 \\
& \ldots & \& & \ldots & \& & 2 \\
& 1 & \& & \ldots & \& & 1 \\
& \ldots & \& & 1 & \& & 1 \\
& \ldots & \& & \ldots & \& & 1 \\
\enddata
\end{deluxetable*}

\begin{figure*} 
\plotone{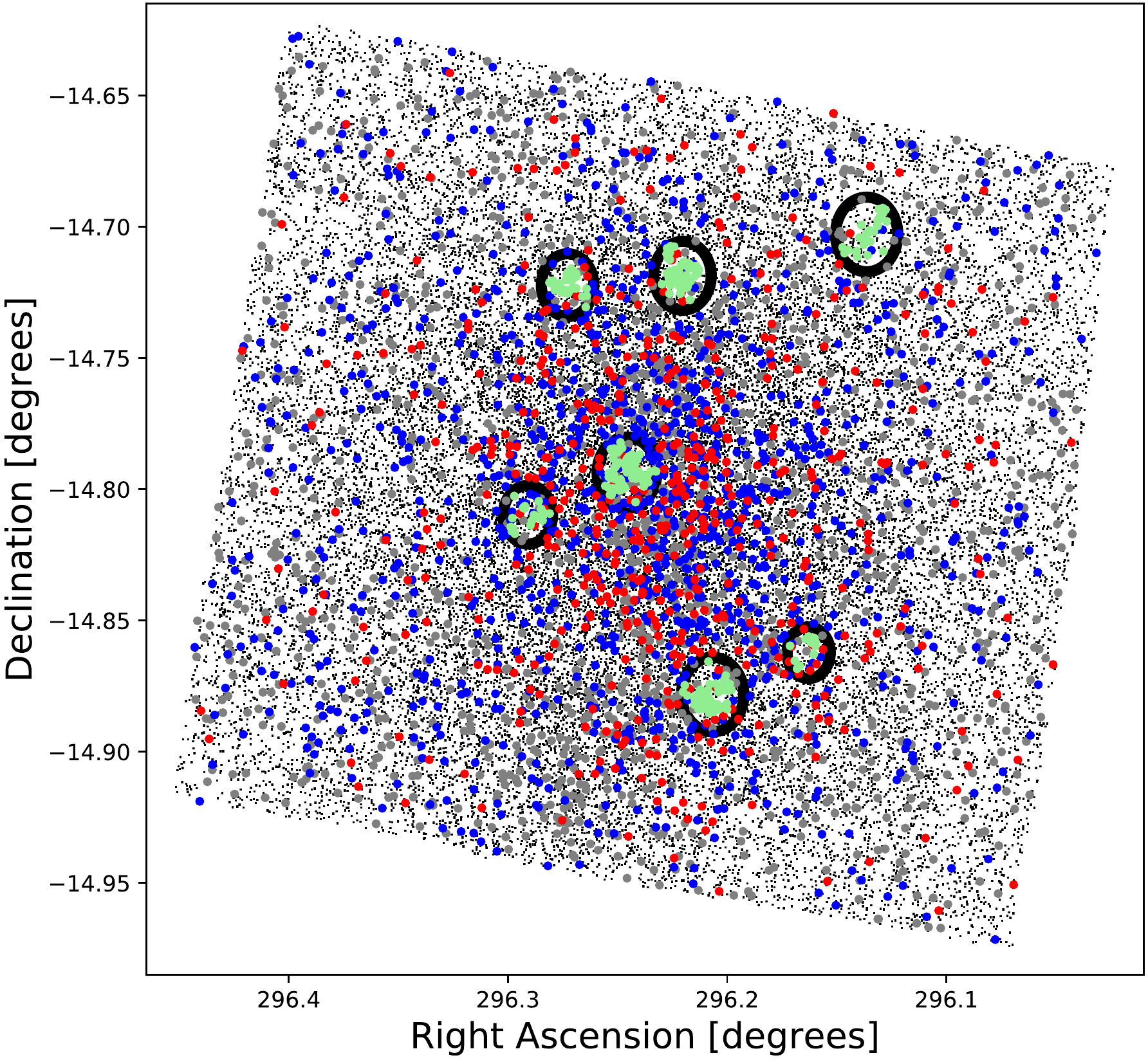}
\caption{
Spatial distribution plot combining the locations of ``RELIABLE" RSGs (gray dots), O-rich AGB stars (blue dots), C-rich AGB stars (red dots), and YSOs (green dots) over all sources included in the master source catalog (black points).
The black rings signify the location of the star-forming regions used in the source classification algorithm detailed in \S3.3.
Prominent and well-known star-forming regions are discernible along the top and bottom of the structure of NGC 6822.
Recent work, presented both here and in \citet{bib:Jones2019}, reveal a young, embedded star-forming region located centrally within the N-S bar.
This cluster, known as Spitzer~I, is located at RA = 19$^{h}$44$^{m}$58.97$^{s}$, Dec = -14$^{d}$47$^{m}$37.39$^{s}$ and presents an exciting new object for future study.
}
\label{fig:posterpic}
\end{figure*}

\begin{figure*} 
\plottwo{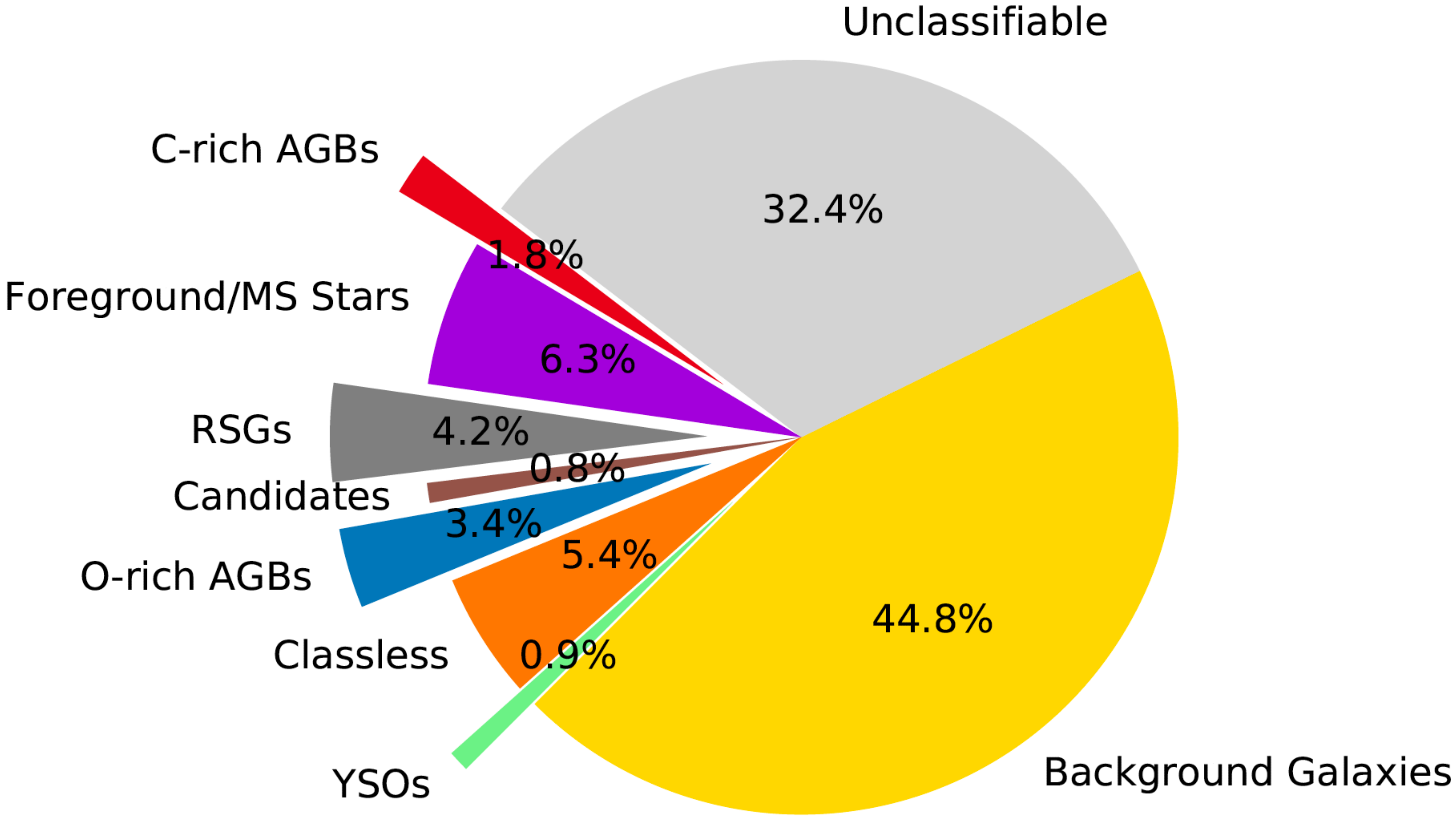}{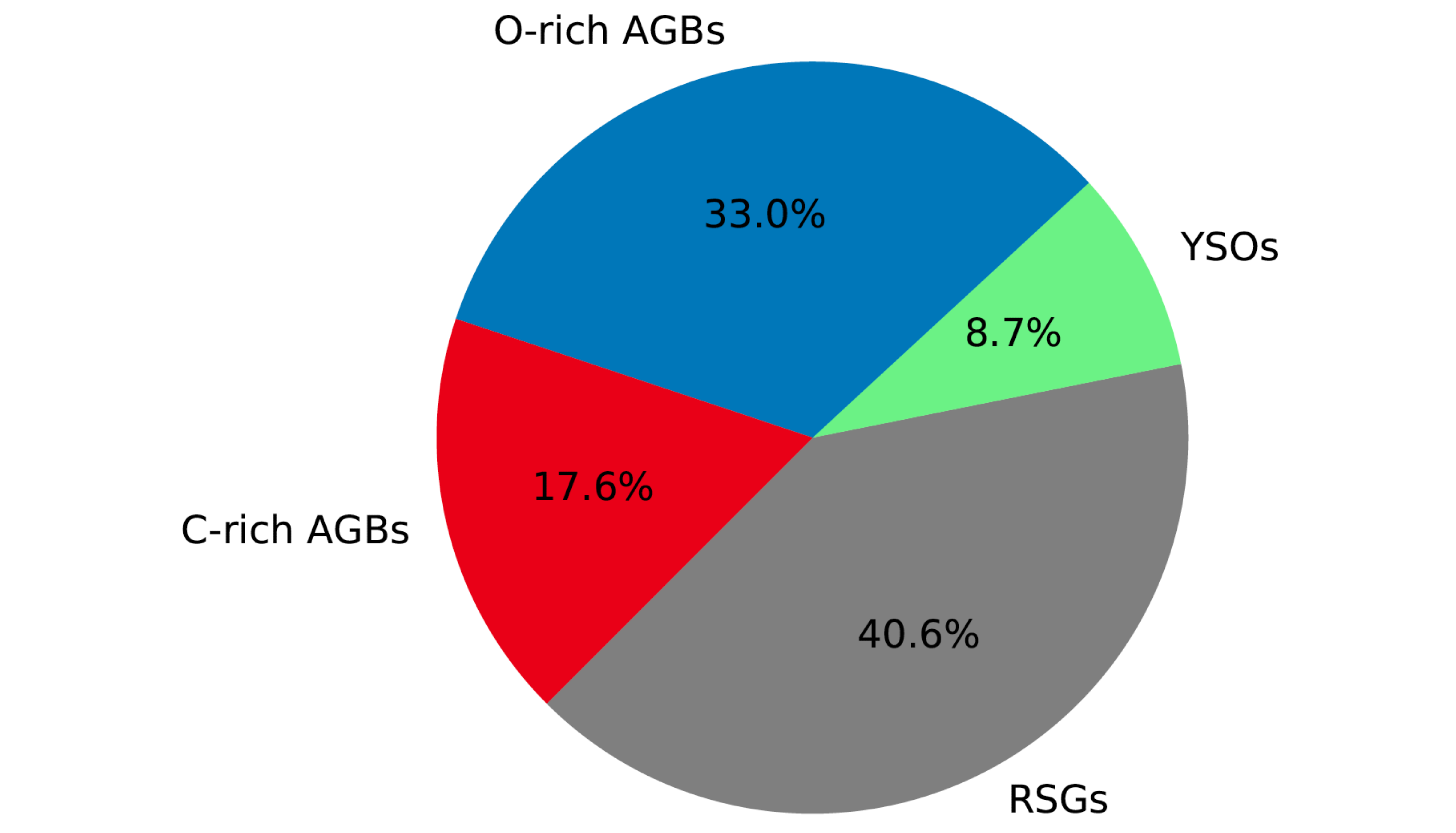}
\caption{
Pie charts illustrating the classification demographics of the entire joined catalog (\emph{left}) and of only the ``RELIABLE" dusty and evolved stars (\emph{right}).
For the ``RELIABLE" catalog, RSGs (gray) number 1,292 sources, O-rich AGB stars (blue) number 1,050 sources, C-rich AGB stars (red) number 560 sources, and YSOs (green) number 277 sources.
All other source types include ``CANDIDATE" objects ($n$ = 260), background galaxies ($n$ = 13,765), foreground/MS stars ($n$ = 1,931), unclassifiable ($n$ = 9,950), and classless ($n$ = 1,660).
}
\label{fig:piecharts}
\end{figure*}

\begin{deluxetable}{ccc} 
\tablenum{10}
\label{tab:catalog_demographics}
\tabletypesize{\small}
\tablewidth{0pt}
\tablecaption{
Demographics of the ``RELIABLE" and ``CANDIDATE" stellar source catalogs, plus all ``OTHER" source types (including background galaxies, foreground/MS stars, unclassifiable, and unclassed).
}
\tablehead{
\colhead{Catalog}&\colhead{Stellar Type}&\colhead{Number of Sources}
}
\startdata
``RELIABLE" & RSGs & 1,292 \\
& O-rich AGB & 1,050 \\
& C-rich AGB & 560 \\
& YSOs & 277 \\
\hline
``CANDIDATE" & RSGs & 65 \\
& O-rich AGB & 59 \\
& C-rich AGB & 27 \\
& Extremely Dusty & 109 \\
\hline
``OTHER" & Background Galaxies & 13,765 \\
& Foreground/MS Stars & 1,931 \\
& Unclassifiable & 9,950 \\
& Classless & 1,660 \\
\enddata
\end{deluxetable}

\subsection{Removing Contaminants} 

\begin{figure*} 
\plotone{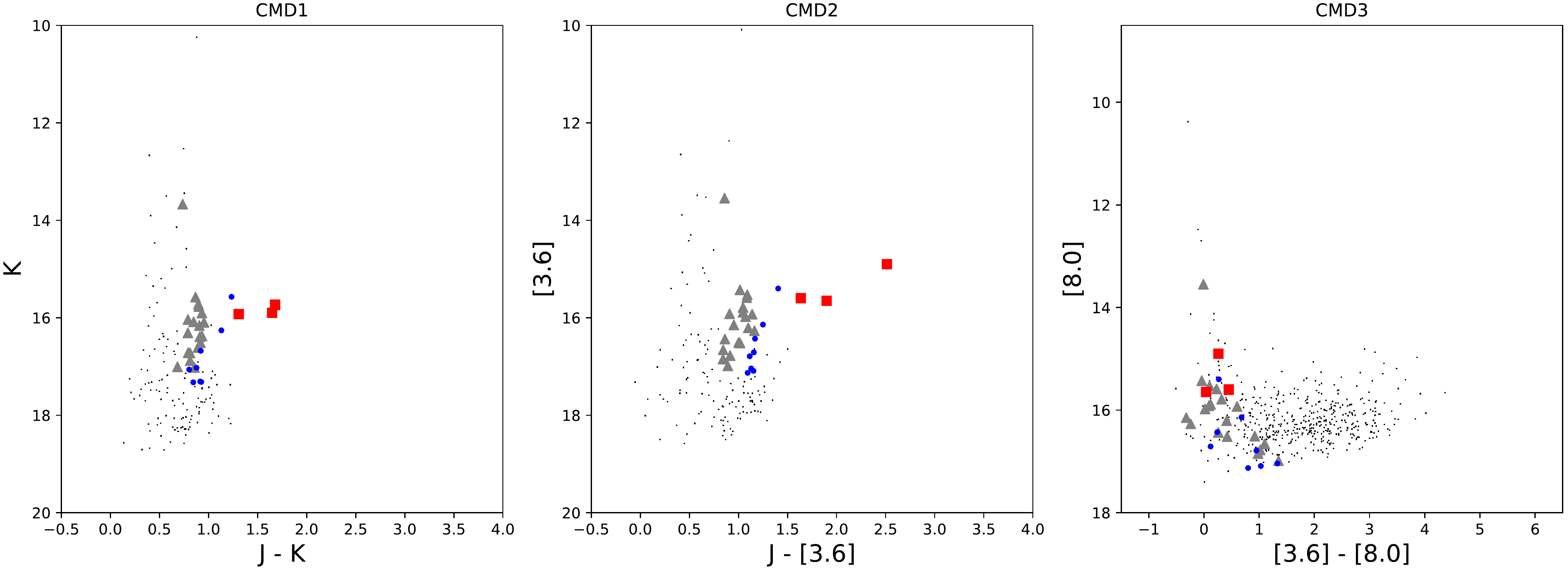}
\caption{
Diagnostic CMDs of sources from the master catalog contained within the off-target comparison box used to compute contamination (gray points; see Tab.\ \ref{tab:contaminants}).
Overplotted are sources from our ``RELIABLE" catalogs, including RSGs (gray triangles), O-rich AGB stars (blue dots), and C-rich AGB stars (red squares).
YSOs are not included here as none appear in the off-target comparison box.
}
\label{fig:OTB}
\end{figure*}

\begin{figure*} 
\plotone{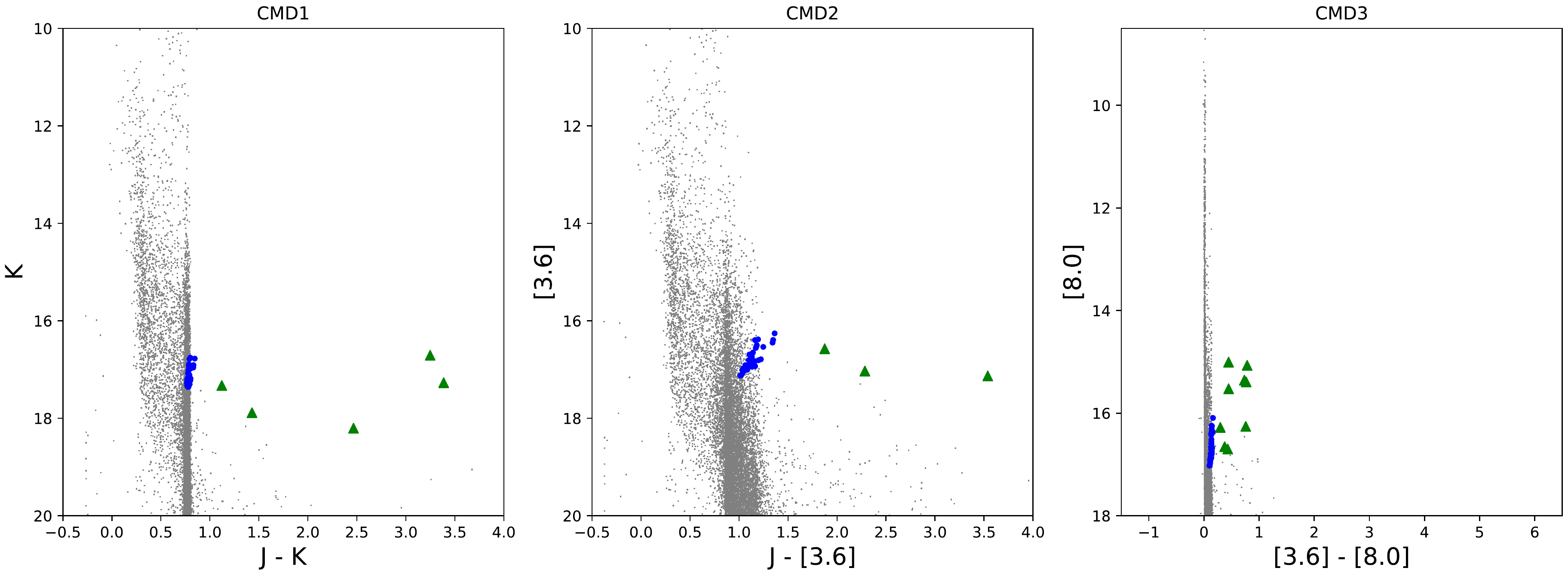}
\caption{
Diagnostic CMDs of sources from from the TRILEGAL Galaxy model, processed through the same color-cut classification criteria as is applied to the ``RELIABLE" sources from our master catalog.
Gray points represent all foreground sources from TRILEGAL.
Model O-rich AGB stars are shown as blue dots, while YSOs are shown as green triangles.
Zero C-rich AGB stars were selected from the modeled data, and so do not appear in these plots.
}
\label{fig:TRILEGAL}
\end{figure*}

\indent Contamination of the various source catalogs developed through this study has been somewhat mitigated via our color-cut specifications, however some number of inadvertent contaminants must be expected.
Near- and mid-IR colors of red supergiant stars (RSGs) are similar to those on the AGB, and at the distance of NGC 6822, unresolved background galaxies, including star-forming galaxies and active galactic nuclei (AGN), have colors which significantly overlap those of dusty sources and YSOs \citep{bib:Jones2019}.
Due to the lower luminosity of the populations, magnitude-based cutoffs used effectively at selecting AGB stars and YSOs in the Magellanic Clouds are less efficient in systems at greater distances such as NGC 6822.
\\
%
%
%
\indent Estimating conservative upper limits for source contamination, we assume a random and homogeneous distribution of background galaxies and foreground stars across our field of interest.
We have adopted an area of 10.80 arcmin$^{2}$ on the outskirts of the galaxy, away from known star-formation regions, to estimate the approximate level of contamination in our source cuts, similar to the method implemented by \citet{bib:Jones2019}.
With 476 point sources in this off-target box identified from the master catalog, this region has a total point-source density of 44.07 sources arcmin$^{2}$.
We calculate the maximum contamination percentage for each of RSG, O-rich AGB star, C-rich AGB star, and dust-enshrouded source candidates (including YSOs) for each of the three diagnostic CMDs (where appropriate), as well as our ``RELIABLE" source lists.
Results are summarized in Table \ref{tab:contaminants}.
From this analysis, we estimate a maximum contamination percentage for our ``RELIABLE" source lists to be 52.93\% for RSGs, 29.52\% for O-rich AGB stars and 20.76\% for C-rich AGB stars.
Because our selection criteria for ``RELIABLE" YSOs requires inclusion within star-forming regions (see Fig.\ \ref{fig:posterpic}), this off-target contamination study necessarily yields zero erroneous sources.
CMDs of the master catalog sources contained within the off-target comparison region are presented as Fig.\ \ref{fig:OTB}.
Overplotted are objects from the ``RELIABLE" catalogs, including RSGs and both O- and C-rich AGB stars.
Note that because no YSOs appear in this off-target comparison region, none appear in these CMDs.
\\
\indent In addition to a statistical investigation of contamination sources, we queried the SIMBAD and VizieR Astronomical Databases using the CDS X-Match Service\footnote{\texttt{http://cdsxmatch.u-strasbg.fr/\#tab=xmatch}} for associations of our RSG, AGB star, and YSO candidates with other known astronomical objects.
Within a 2$\arcsec$ radius, 646 RSGs, 840 O-rich AGB stars, 530 C-rich AGB stars, and 148 YSOs were matched to sources with literature classifications.
These are mostly comprised of AGB star candidates identified using optical or near-IR colors (\citealp{bib:Sibbons2012, bib:Whitelock2013, bib:Kang2006, bib:Letarte2002}).
In addition, there are matched classifications indicative of youth, e.g., \HA-emitting objects \citep{bib:Massey2007}, CO-bright clumps \citep{bib:Schruba2017}, compact \HII\ regions \citep{bib:Hernandez-Martinez2009a}, and presence within a star-forming region \citep{bib:Melena2009}.
Finally, a handful sources have been matched to spectroscopically-confirmed red giant stars \citep{bib:Kirby2017}.
\\
\indent Of the RSGs, three have been matched to red giants, four have been matched to \HA\ emission-line stars, and 23 to star-forming regions.
Twenty-three sources had conflicting classifications (all but three of which are classified as both star-forming regions and AGB stars).
For O-rich AGB stars, two have been matched to red giants, four have been matched to \HA\ emission-line stars, 15 to star-forming regions, and one to a CO-bright clump.
Conflicting classifications occurred for 21 sources.
C-rich AGB stars match to three red giants, five \HA\ emission-line stars, nine star-forming regions, and 2 CO-bright clumps.
Seventeen sources presented conflicting classifications.
Finally, one YSO was matched to a red giant (likely the result of a chance superposition of one optically-bright and one IR-bright source in the crowded field), seven matched to \HII\ regions, two to planetary nebulae (PNe), 20 to \HA\ emission-line stars, to 10 star-forming regions, 10 to \HA\ emission regions from UV data, and 80 to CO-bright clumps.
We note that mid-IR \emph{Spitzer} colors for YSOs and PNe are indistinguishable and require spectral disentanglement and are thus potentially significant sources of contamination \citep{bib:Jones2017a}.
There were 31 instances of conflicting classifications.
\\
\indent Finally, we employed the TRILEGAL service\footnote{\texttt{http://stev.oapd.inaf.it/cgi-bin/trilegal}} to better understand foreground sources of contamination.
This tri-dimensional model of the Galaxy \citep{bib:Girardi2012} is used to populate the field of view to NGC 6822 with artificial foreground stars, which are then run through the same color-cut classification criteria as is applied to our master catalog sources.
Because completeness of the data is not well defined at blue wavelengths, this technique was implemented to grasp the likelihood of foreground contamination by O- and C-rich AGB stars and YSOs.
CMDs of all sources obtained from the TRILEGAL Galaxy model are presented as Fig.\ \ref{fig:TRILEGAL}.
This data was subsequently processed through the same color-cut classification criteria as applied to the ``RELIABLE" dusty and evolved star catalogs, producing artificial contaminants which have been overplotted.
In total, we found 41 simulated O-rich AGB star contaminants, zero such C-rich AGB stars, and nine YSOs.
%

\begin{deluxetable*}{ccccccc} 
\tablenum{11}
\label{tab:contaminants}
\rotate
\tabletypesize{\normalsize}
\tablewidth{0pt}
\tablecaption{Contamination estimates from off-target comparison region.}
\tablehead{
\colhead{CMD}&\colhead{Source}&\colhead{Number}&\colhead{Contaminant}&\colhead{Average Number}&\colhead{Number of Color-}&\colhead{Maximum}
\\
\colhead{Catalog}&\colhead{Type}&\colhead{of Sources}&\colhead{Sources}&\colhead{of Contaminants}&\colhead{-Selected Sources}&\colhead{Contamination}
\\
\colhead{}&\colhead{}&\colhead{in Box}&\colhead{[arcmin$^{-2}$]}&\colhead{in Entire FoV}&\colhead{in Entire FoV}&\colhead{Percentage [\%]}
}
\startdata
$K$ vs.\ $J-K$ & RSGs & 40 & 3.70 & 1550.00 & 2075 & 74.70 \\
$K$ vs.\ $J-K$ & O-rich AGB & 7 & 0.65 & 271.25 & 1002 & 27.07 \\
$K$ vs.\ $J-K$ & C-rich AGB & 7 & 0.65 & 271.25 & 907 & 29.91 \\
$K$ vs.\ $J-K$ & Dust-enshrouded & 0 & 0.00 & 0.00 & 144 & 0.00 \\
{[}3.6] vs.\ $J-[3.6]$ & RSGs & 19 & 1.76 & 736.25 & 1571 & 46.87 \\
{[}3.6] vs.\ $J-[3.6]$ & O-rich AGB & 15 & 1.39 & 581.25 & 1545 & 37.62 \\
{[}3.6] vs.\ $J-[3.6]$ & C-rich AGB & 3 & 0.28 & 1916.25 & 713 & 16.30 \\
{[}3.6] vs.\ $J-[3.6]$ & Dust-enshrouded & 0 & 0.00 & 0.00 & 230 & 0.00 \\
{[}8.0] vs.\ $[3.6]-[8.0]$ & M-type (all) & 3 & 0.28 & 116.25 & 674 & 17.25 \\
{[}8.0] vs.\ $[3.6]-[8.0]$ & C-rich AGB & 33 & 3.06 & 1278.75 & 2949 & 43.36 \\
{[}8.0] vs.\ $[3.6]-[8.0]$ & Dust-enshrouded & 3 & 0.28 & 116.25 & 630 & 18.45 \\
``RELIABLE" & RSGs & 19 & 1.76 & 736.25 & 1391 & 52.93 \\
``RELIABLE" & O-rich AGB & 8 & 0.74 & 310.00 & 1050 & 29.52 \\
``RELIABLE" & C-rich AGB & 3 & 0.28 & 116.25 & 560 & 20.76 \\
``RELIABLE" & YSOs & 0 & 0.00 & 0.00 & 277 & 0.00 \\
\enddata
\end{deluxetable*}

\section{Discussion} 

\begin{figure*} 
\plotone{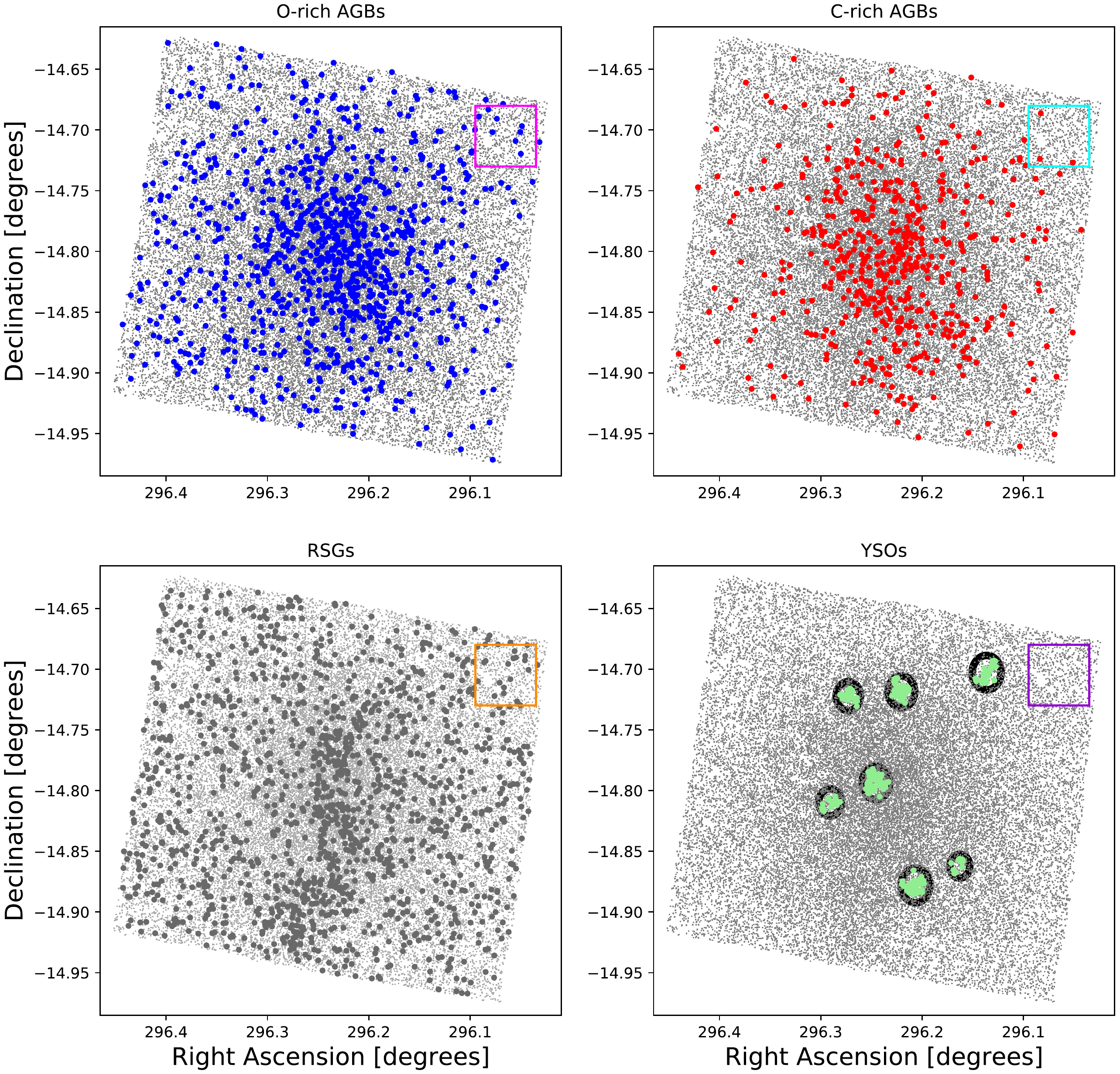}
\caption{
Spatial distribution plots of O-rich AGB star (blue points; \emph{upper-left}), C-rich AGB star (red points; \emph{upper-right}), RSG (gray points; \emph{lower-left}), and YSO candidates (green points within black rings; \emph{lower-right}).
Colored points are overplotted on all sources in the NGC 6822 master catalog (gray dots).
The region used to determine contamination percentages (presented in Table \ref{tab:contaminants}) is included in the top right of each plot as a colored box.
}
\label{fig:spatdist_DESALL}
\end{figure*}

\indent With the categorizations of the stellar types established in \S3.3., we now examine the spatial distribution of dusty and evolved stars in NGC 6822 and compare these results with other studies available in the literature.
A plot of all points in RA and Dec space from our master catalog effectively produces a synthetic map of the galaxy.
A comparison with the image of NGC 6822 (Fig.\ \ref{fig:NGC6822}) reveals that such a spatial distribution plot features the same main structural components, including the central bar (running N-S) within which the density of the individual sources is highest.
We present the spatial distributions of the evolved star populations from our ``RELIABLE" catalogs as Fig.\ \ref{fig:spatdist_DESALL}.
The AGB star candidates (O-rich, \emph{upper-left}; C-rich, \emph{upper-right}) appear correlated with the extent and structure of the galaxy.
These points are situated primarily within the central N-S bar of NGC 6822, and therefore appear to have been successfully selected by the color-cut classification technique to be physically associated with the galaxy itself.
The RSG candidates (\emph{lower-left}) demonstrate a strong clustering along the central bar as well, though with a higher number of points distributed homogeneously across the field of view.
As discussed in \S3.4, the RSG color cut is strongly influenced by contamination from foreground sources, which may account for this effect.
Finally, the selection criteria for YSO candidates (\emph{lower-right}) necessitates that they are confined to within the seven major star-formation regions as defined in \S3.3.
These include the well-known Hubble star-forming regions (i.e., Hubble X, Hubble V, and Hubble 1 and III along the north side, and Hubble IV on the south side of NGC 6822), as well as three newly-identified, young, embedded clusters dubbed Spitzer I, II, and III, which were introduced in \citet{bib:Jones2019} and discussed further in \S4.2.

\subsection{Calculations of Stellar Metallicity} 

\begin{figure*} 
\plotone{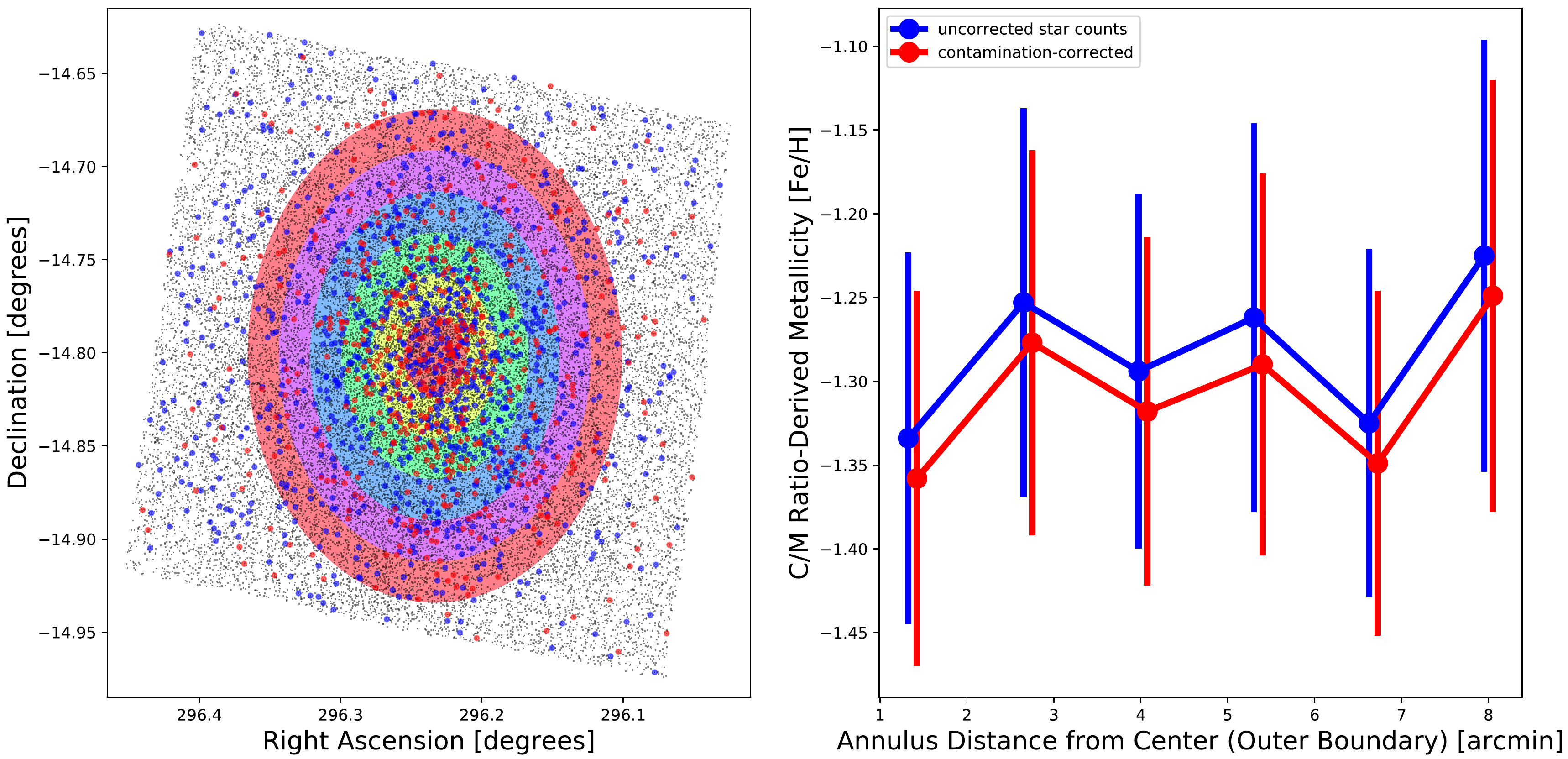}
\caption{
Metallicity analysis via C/M ratio in NGC 6822 along circular annuli originating at the center and increasing outwards (\emph{left}).
O-rich AGB stars are represented as blue dots, while C-rich AGB stars are represented as red dots.
These overplot all sources from the master catalog, shown as gray dots.
Each annulus is 1.325 arcsec in width, corresponding to the quarter-light (red ellipse), half-light (yellow ring), three-quarter-light (green ring), full-light (blue ring), one-and-one-quarter-light (violet ring), and one-and-one-half-light (orange ring) radii.
Stellar metallicity [Fe/H] computed by the C/M ratio relation of \citet{bib:Cioni2009} is plotted as a function of annulus distance from the galaxy center (\emph{right}; see Table \ref{tab:radial_metallicity}).
Values from uncorrected star counts are presented in blue, while contamination-corrected values are presented in red (with a small arbitrary $x$-axis value offset to avoid overlap).
We find an increase in metallicity with increasing distance.
The rate of increase is very slight, however, and all values of metallicity agree within their respective errors, therefore giving only a tenuous suggestion of a radial metallicity gradient.
}
\label{fig:C-to-O_twofer}
\end{figure*}

\indent With a census of ``RELIABLE" O- and C-rich AGB star candidates determined through our color-cut classifications, we may investigate the stellar metallicity characteristics of this galaxy.
The ratio of the number of C-rich (C-type) to O-rich (M-type) AGB stars can be used as a gauge of the star formation environment.
For a lower initial ambient metallicity, fewer dredge-up events are necessary to form a C-rich atmosphere, and thus a higher measured value of the overall C/M ratio is expected compared to a solar metallicity environment (\citealp{bib:Groenewegen2006} (and references therein), \citealp{bib:Sibbons2012, bib:Boyer2017, bib:Boyer2019}).
In addition, a lower metallicity environment will push the AGB evolutionary track toward higher temperatures.
This has the further result of reducing the numbers of M-type stars while simultaneously increasing the number of K spectral type RGB stars \citep{bib:Sibbons2012, bib:Marigo1999, bib:IbenRenzini1983}.
By the relation of \citet{bib:BattinelliDemers2005} and refined by \citet{bib:Cioni2009},
\small
\renewcommand{\arraystretch}{0.85}
\begin{eqnarray}
\label{eq:Cioni_metallicity}
\begin{array}{r}
{\rm [Fe/H]} \ = \ -1.39 \pm 0.06 - 0.47 \pm 0.10 \ \times\ {\rm log(C/M)} \\
\end{array}     
\end{eqnarray}
\normalsize
The C/M ratio for NGC 6822, computed over the entire field of view for our study, is 0.533 $\pm$ 0.028.
We therefore find a metallicity for NGC 6822 of [Fe/H] = -1.262 $\pm$ 0.098.
Corrected for maximum source contamination (see Table \ref{tab:contaminants}), the C/M ratio becomes 0.599 $\pm$ 0.036, which gives a metallicity [Fe/H] = -1.286 $\pm$ 0.095.
The study of \citet{bib:Sibbons2012} report a C/M ratio of 0.23 $\pm$ 0.01 over their full-observed area, which produces a metallicity estimate of [Fe/H] = -1.14 $\pm$ 0.08 utilizing the same \citet{bib:Cioni2009} relation.
Within 4 kpc of the center, however, they report a C/M ratio of 0.48 $\pm$ 0.02, equivalent to [Fe/H] = -1.24 $\pm$ 0.07.
For a distance to NGC 6822 of 490 kpc, a radius 4 kpc from the galaxy center is equivalent to 28.06$\arcmin$, which is larger than the field of view of our study as dictated by the extent of the \emph{Spitzer} data.
This second metallicity from \citet{bib:Sibbons2012} is therefore a much more appropriate comparison to our own reported value, which agree within the respective errors.
\\
\indent In addition, we have investigated the AGB stars in NGC 6822 for possible structure in the metallicity distribution of this galaxy.
First, by segregating the AGB stars along the central N-S bar, we investigate the variability of metallicity across the galactic axis of rotation.
We adopt the coordinates for the center of NGC 6822 from \citet{bib:Sibbons2012} as RA = 19$^{h}$44$^{m}$56$^{s}$, Dec = -14$^{d}$48$^{m}$06$^{s}$, and the angle of the axis rotation of 330$^{\circ}$ adopted from \citet{bib:McConnachie2012}.
On the left side of NGC 6822, we find a C/M ratio of 0.507 $\pm$ 0.038, which corresponds by Eq.\ \ref{eq:Cioni_metallicity} to [Fe/H] = -1.251 $\pm$ 0.105.
On the right side of the galaxy, we find a C/M ratio of 0.561 $\pm$ 0.041, giving a metallicity of [Fe/H] = -1.272 $\pm$ 0.100.
When corrected for maximum source contamination, the left-side C/M ratio becomes 0.570 $\pm$ 0.049, corresponding to a metallicity of [Fe/H] = -1.275 $\pm$ 0.102.
The contamination-corrected values for the right side are a C/M ratio of 0.631 $\pm$ 0.053, which gives a metallicity [Fe/H] = -1.296 $\pm$ 0.097.
The measured stellar abundances of these two halves of NGC 6822 fall neatly to either side of the overall metallicity of [Fe/H] = -1.262 $\pm$ 0.098 ([Fe/H] = -1.286 $\pm$ 0.095 when corrected for contamination) discussed earlier, however the magnitude of the difference is minuscule in comparison with the respective errors.
We therefore conclude that the rotation of NGC 6822 has no impact on the measured stellar metallicity of the system.
\\
\indent Second, we have isolated the C- and M-type AGB star populations in six concentric rings of width equal to the quarter-light radius (1.325$\arcmin$; $\sim$0.189 kpc), initiating from the center of the galaxy.
Within these rings, we compute the C/M ratio and the associated value of [Fe/H] out to a distance of 7.95$\arcmin$ ($\sim$1.133 kpc), equivalent to the one-and-one-half-light radius.
We find a minimum value of [Fe/H] = -1.334 $\pm$ 0.111 in the most central annulus, with metallicities increasing with expanding radius to a maximum of [Fe/H] = -1.225 $\pm$ 0.129 in the outermost ring ([Fe/H] = -1.358 $\pm$ 0.112 to -1.249 $\pm$ 0.129 when contamination corrected).
These results are illustrated in Fig.\ \ref{fig:C-to-O_twofer} and summarized in Table \ref{tab:radial_metallicity}.
The rate of increase we find for NGC 6822 is quite minimal, however, and all values of metallicity agree within their respective errors, therefore giving only a very slight suggestion of a radial metallicity gradient.
We therefore do not suggest the reality of a radial metallicity gradient in NGC 6822.
\\
\indent Overall, these values of the stellar metallicity depend on the implicit assumption that all C- and O-rich AGB star candidates from our ``RELIABLE" catalog are actually AGB stars.
Computations utilizing maximum contamination percentage corrected values are only avenue in this analysis.
While our classification methodology requires candidacy from multiple CMDs to be marked as ``RELIABLE" for the sake of redundancy, the presence of contaminant RGBs in our M-type star catalog cannot be ignored.
Our C/M ratios and [Fe/H] values must therefore be considered as lower limits.
The true nature of these sources requires more sophisticated analysis than color-cut segregations can provide.
Future studies employing SED fitting promise to refine this value even further with higher-confidence identifications of the relevant stellar populations.

\begin{deluxetable}{cccccc} 
\tablenum{12}
\label{tab:radial_metallicity}
\rotate
\tabletypesize{\normalsize}
\tablewidth{0pt}
\tablecaption{
C/M ratio and [Fe/H] values for concentric radial rings originating from the center of NGC 6822.
}
\tablehead{
\colhead{Inner Ring Boundary}&\colhead{Outer Ring Boundary}&\colhead{C/M}&\colhead{[Fe/H]}&\colhead{C/M}&\colhead{[Fe/H]}
\\
\colhead{Distance from Center}&\colhead{Distance from Center}&\colhead{(uncorrected}&\colhead{(uncorrected}&\colhead{(contamination}&\colhead{(contamination}
\\
\colhead{[arcmin]}&\colhead{[arcmin]}&\colhead{star counts)}&\colhead{star counts)}&\colhead{-corrected)}&\colhead{-corrected)}
}
\startdata
0.000 & 1.325 & 0.761 $\pm$ 0.141 & -1.334 $\pm$ 0.111 & 0.856 $\pm$ 0.183 & -1.358 $\pm$ 0.112 \\
1.325 & 2.650 & 0.511 $\pm$ 0.065 & -1.253 $\pm$ 0.116 & 0.575 $\pm$ 0.084 & -1.277 $\pm$ 0.115 \\
2.650 & 3.975 & 0.625 $\pm$ 0.076 & -1.294 $\pm$ 0.106 & 0.703 $\pm$ 0.098 & -1.318 $\pm$ 0.104 \\
3.975 & 5.300 & 0.544 $\pm$ 0.073 & -1.262 $\pm$ 0.116 & 0.612 $\pm$ 0.094 & -1.290 $\pm$ 0.114 \\
5.300 & 6.625 & 0.726 $\pm$ 0.104 & -1.325 $\pm$ 0.104 & 0.817 $\pm$ 0.134 & -1.349 $\pm$ 0.103 \\
6.625 & 7.950 & 0.446 $\pm$ 0.073 & -1.225 $\pm$ 0.129 & 0.502 $\pm$ 0.094 & -1.249 $\pm$ 0.129 \\
\enddata
\end{deluxetable}

\subsection{The Young, Embedded Star-Forming Regions Spitzer I, Spitzer II, and Spitzer III} 

\indent The bright, well-known star-forming complexes of NGC 6822 (e.g., Hubble X, Hubble V, and Hubble I and III situated along the north side, and Hubble IV situated on the south side of the galaxy), distinguished by optically-conspicuous \HA\ emission (see Fig.\ \ref{fig:NGC6822}), have been studied extensively across several wavelength regimes.
\citet{bib:Hubble1925} first presented optical plates of NGC 6822, identifying conspicuous star-forming regions.
\HA\ fluxes of the two brightest \HII\ regions, Hubble V and Hubble X, were obtained by \citet{bib:Kennicutt1979}, which were later observed with \hst\ by \citet{bib:ODell1999}.
\citet{bib:Hodge1988} compiled an atlas of \HII\ regions via optical survey, while \citet{bib:Hodge1989} provided an \HA\ luminosity function and \HII\ region size distribution.
\citet{bib:Gallagher1991} and \citet{bib:Israel1996} studied the star-forming regions in the far-IR.
CO emission of was studied to trace molecular gas by \citet{bib:Israel2003} and later, on a larger scale, by \citet{bib:Schruba2017}.
\citet{bib:Cannon2006} presented \emph{Spitzer} imaging of emission regions in NGC 6822 to study the nature of IR, \HA, \HI, and radio continuum emission.
\citet{bib:Gouliermis2010} employed optical photometry and stellar density maps to show the hierarchical stellar structure in the galaxy, linking some stellar concentrations with IR-bright complexes identified with \emph{Spitzer} imaging from \citet{bib:Cannon2006}.
Finally, several studies have used the massive \HII\ regions of NGC 6822 as specimens of extragalactic star formation, employing \emph{Spitzer} spectroscopy to study photodissociation, fine-structure lines, and abundance ratios (e.g., \citealp{bib:Lee2005a, bib:HunterKaufman2007, bib:Rubin2016}).
\\
\indent The spatial distribution of ``RELIABLE" YSO candidates is illustrated in the lower-right of Fig.\ \ref{fig:spatdist_DESALL}.
Sources are tightly clustered in compact regions consistent with the well-known star-forming regions of NGC 6822.
In addition to these, we have identified comparable numbers of dusty sources grouped in other tight clusters.
Located coincident with the central N-S bar, these collections of red-excess objects appear to be massive, embedded regions of star formation that have been otherwise hidden from previous optical surveys.
Only via concentrated effort with near- and mid-IR photometry are these obscured clusters of YSO candidates revealed.
The most central star-forming region, dubbed Spitzer I by \citet{bib:Jones2019}, is located at RA = 19$^{h}$44$^{m}$58.97$^{s}$, Dec = -14$^{d}$47$^{m}$37.39$^{s}$ and is home to 80 of the ``RELIABLE" YSOs identified in this study.
\citet{bib:Cannon2006} categorized this region as the infrared source ``NGC 6822 11", cross-listed as Hubble VI and VII, while \citet{bib:Gratier2010} studied the molecular cloud complexes of this region in CO.
Immediately adjacent to Spitzer I is the smaller Spitzer II, located at RA = 19$^{h}$45$^{m}$09.80$^{s}$, Dec = -14$^{d}$48$^{m}$34.77$^{s}$, and within which we have found 18 YSOs.
Finally, next to Hubble IV in the south of NGC 6822 is Spitzer III, located at RA = 19$^{h}$44$^{m}$39.08$^{s}$, Dec = -14$^{d}$51$^{m}$44.12$^{s}$, and which contains 14 YSOs.
\\
\indent The high concentration of YSO candidates within these star-forming regions, along with the strong likelihood of source confusion due to crowding and resolution effects, suggests a larger number of individual YSOs than our selection criteria are capable of singling out.
The observational signature of Spitzer I may in fact indicate the presence of a proto-super star cluster (SSC;\ $M_{*}$ $\gtrsim$ 10$^{5}$ M$_{\odot}$).
SSCs are representative of a star formation mode where stellar surface densities exceed that of \HII\ regions and OB associations by orders of magnitude \citep{bib:Nayak2019}, and may be early-Universe analogs to modern-day globular clusters.
The most well-known SSC is the cluster R136, which resides within the massive LMC star-forming region 30 Doradus.
A parallel star-forming complex known as N79, identified to possess \HA-emitting sources first by \citet{bib:Henize1956}, occupies space symmetrically across the bar of the LMC, discernible in the distribution of \HI\ gas (see Fig.\ 1 of \citealp{bib:Ochsendorf2017}).
Within the southernmost of three giant molecular clouds (GMCs) in N79 is the object H72.97-69.39, originally thought to be the most luminous YSO in the entire LMC \citep{bib:Seale2014}, but recently identified as a proto-SSC \citep{bib:Ochsendorf2017, bib:Nayak2019}.
\\
\indent In contrast the optically-luminous and -conspicuous 30 Doradus, the N79 region is weak in optical star-formation tracers such as \HA\ emission.
\citet{bib:Ochsendorf2017} have found N79 to be extremely active, however, hosting a densely-packed population of YSOs.
Similarly, while roughly equivalent in physical size to Hubble V, the brightest \HII\ region in NGC 6822 (radius $\approx$ 120 pc), our identification of 80 YSOs associated with Spitzer I is $\sim$50\% more than were found in Hubble V ($n$ = 55), demonstrating a high density of young stars.
Spitzer I is rich in molecular gas (inferred from 8 $\mu$m emission;\ \citealp{bib:Sandstrom2012}), and is probably the youngest star-forming region in NGC 6822 \citep{bib:Jones2019}.
Furthermore, analyses of star-formation tracers reveal a higher level of IR flux as compared to UV or \HA, providing a strong indication that the SFR is expected to increase.
Similarly to the LMC object H72.97-69.39 in N79, Spitzer I has yet to reach its peak level of star-formation activity, and may potentially become an SSC.
While the question of exactly \emph{how} SSCs form remains a mystery, future detailed studies of H72.97-69.39 and Spitzer I may shed light on this extreme star-formation phenomenon.
\\
\indent The recent work of \citet{bib:Jones2019} employed the joined master catalog developed in this study to specifically characterize the YSO population of NGC 6822.
Utilizing an independent set of color-cut criteria, \citet{bib:Jones2019} have identified 90 YSOs within the region of space affiliated with Spitzer I.
Their classification system further employed spectral energy distribution (SED) modeling, finding the Spitzer I YSOs exhibiting properties characteristic of the earliest stages of formation.
It is rich in molecular gas, and has substantial emission in 8 $\mu$m as well as some emission in 24 $\mu$m, but is faint in \HA\ and UV.
Combined with a large \HI\ mass, the large population of high-confidence YSO detections strongly indicate that this star-forming region is young and very embedded, exhibiting the highest SFR known in NGC 6822 \citep{bib:Jones2019}.
Future observations with \jwst\ promise to characterize this potential proto-SSC with exceptional detail.

%
%
%

\section{Summary} 

\indent In this study, we have combined near- and mid-IR photometric data from \citet{bib:Sibbons2012} and \citet{bib:Khan2015} of the nearby, metal-poor dwarf star-forming galaxy NGC 6822 to produce a catalog useful for characterization of dusty and evolved stars.
Through implementation of KDE local minima analyses, including the Monte Carlo method KDE (or ``MCKDE") technique used to define the TRGB, we have produced color cuts to categorize sources into several stellar type categories, including RSGs, O-rich AGB star candidates, C-rich AGB star candidates, and dust-enshrouded sources (including YSOs).
These color cut classifications were developed for three diagnostic CMDs, which span a wide range in wavelength space afforded by the available data, and include $K$ vs.\ $J-K$ (CMD1), [3.6] vs.\ $J$ -- [3.6] (CMD2), and [8.0] vs.\ [3.6] -- [8.0] (CMD3).
\\
\indent From these KDE-derived color cut categorizations, overall source type classifications were established to produce ``RELIABLE" and ``CANDIDATE" catalogs.
From a catalog with $N$ = 30,745 total sources, $n$ = 3,179 have been classified as ``RELIABLE" dusty and evolved stars (10.3\% of the total catalog), while $n$ = 260 have been classified as ``CANDIDATE" (0.8\% of the total catalog).
Of these ``RELIABLE" sources, $n$ = 1,292 are categorized as RSGs, $n$ = 1,050 are categorized as O-rich AGB stars, $n$ = 560 are categorized as C-rich AGB stars, and $n$ = 277 are categorized as YSOs.
The small number of ``CANDIDATE" sources include $n$ = 65 RSGs, $n$ = 59 O-rich AGB stars, $n$ = 27 C-rich AGB stars, and $n$ = 109 dust-enshrouded sources (which may include AGB stars or YSOs).
The combined CMD color-cut categorizations of these objects are more ambiguous than for our ``RELIABLE" object classifications.
We categorize the majority of the remainder sources as either background galaxies ($n$ = 13,765), foreground and MS stars ($n$ = 1,931), or, because they do not fall into type categorizations in any of the three diagnostic CMDs and/or possess only upper limits to their photometric uncertainty values, are considered unclassifiable ($n$ = 9,950).
Finally, the remaining $n$ = 1,660 sources do not present an obvious type from our analysis and are therefore considered ``CLASSLESS".
\\
\indent The spatial distribution of the ``RELIABLE" evolved stars shows strong coincidence to the central N-S bar of NGC 6822 in the case of O- and C-rich AGB star candidates.
Employing the relation of \citet{bib:Cioni2009}, we determine a stellar metallicity of NGC 6822 of [Fe/H] = -1.286 $\pm$ 0.095 as defined by the ratio of C- to M-type AGB stars.
This value is consistent with other studies in the literature.
Major star-forming regions of NGC 6822 are heavily permeated with ``RELIABLE" YSO candidates.
The results of this study, in concert with those presented in \citet{bib:Jones2019}, imply the existence of three young, embedded star-forming clusters in the central region of NGC 6822 that have remained relatively unexplored.
One such cluster, known as Spitzer I, is home to a greater number of YSO candidates than the galaxy's bright, well-known star-forming regions such as Hubble V.
In this regard, Spitzer I may represent a proto-super star cluster in the early stages of its evolution, sharing interesting characteristics with the proto-SSC H72.97-69.39 found in the N79 region of the Large Magellanic Cloud \citep{bib:Ochsendorf2017, bib:Nayak2019} and necessitating detailed followup study with dedicated observing time.
\\
\indent This project was developed in effort to better understand the role that dust plays in star-forming systems with metal-poor environmental conditions analogous to those which populated the early Universe.
While the bulk of the enrichment history of the Universe is traced to this epoch of peak star formation ($z$ $\sim$ 1.5--2), within these galaxies the low- to intermediate-mass MS stars would not have had sufficient time to reach the AGB phase of their evolution.
As AGB stars are a primary producer of dust, a census of these evolved stars and study of their enrichment properties in extremely metal-poor environments is critical to understanding the early Universe.
\\
\indent Future \jwst\ GTO programs investigating the dusty and evolved-star populations of metal-poor, early-Universe analog systems will deliver high-quality observational data of NGC 6822 and the blue compact dwarf (BCD) galaxy I Zw 18.
The near- and mid-IR wavelength coverage of the Near-Infrared Camera (NIRCam) and Mid-Infrared Instrument (MIRI) will provide photometry similar to those used in this study.
Development of best practices are therefore of great importance for the upcoming \jwst\ era, in which quick turnaround for spectral followup of interesting sources will be necessary.
\\
\\
\\
\noindent \textbf{Acknowledgements:}\
The authors would like to thank the referee for the useful comments provided which helped to improve this paper.
We thank Bernie Shiao for his assistance running CASJobs queries, which were used in developing the joined master catalog of photometric sources.
We additionally extend our thanks to Owen Boberg and Peter Scicluna for their help with developing Python routines used in the source classifications.
ASH and MM acknowledge support from NASA grant NNX14AN06G.
LG was funded via the Space Astronomy Summer Program (SASP) at STScI.
Thanks to William Paranzino, a Johns Hopkins University summer intern, for his help in the early assessment of contaminants to the catalog.
OCJ has received funding from the EUs Horizon 2020 programme under the Marie Sklodowska-Curie grant agreement No 665593 awarded to the STFC.
This research made use of Astropy\footnote{\texttt{http://www.astropy.org/}}, a community-developed core Python package for Astronomy \citep{bib:Robitaille2013};
APLpy, an open-source plotting package for Python \citep{bib:RobitailleBressert2012};
and the SIMBAD database, operated at CDS, Strasbourg, France \citep{bib:Wenger2000}.


\end{document}